\documentclass[12pt,twoside]{article}
\usepackage{amsmath}

\newif\ifdouble\doubletrue \doublefalse

\DeclareMathAlphabet\mathfrak{U}{euf}{m}{n}
\DeclareFontFamily{U}{euf}{}
\DeclareFontShape{U}{euf}{m}{n}{
  <5> <6> <7> <8> <9> gen * eufm
  <10> <10.95> <12> <14.4> <17.28> <20.74> <24.88> eufm10
  }{}
\DeclareMathAlphabet\mathbb  {U}{msb}{m}{n}
\DeclareFontFamily{U}{msb}{}
\DeclareFontShape{U}{msb}{m}{n}{
  <5> <6> <7> <8> <9> gen * msbm
  <10> <10.95> <12> <14.4> <17.28> <20.74> <24.88> msbm10
  }{}
 
\def\one{\mathsf{1}}
\def\iu{\mathrm{i}}
\def\dslash{\partial \kern-0.52em \raisebox{0.1em}{/} \kern 0.1em}

\allowdisplaybreaks[2]
\ifdouble
  
\else
\fi

\makeatletter
\def\section{\@startsection{section}{1}{\z@}{-3.25ex plus -1ex minus -.2ex}%
            {1.5ex plus .2ex}{\normalfont\bfseries}} 

\renewenvironment{thebibliography}[1]
        {\frenchspacing\small
         \begin{list}{\arabic{enumi}.}
        {\usecounter{enumi}\setlength{\parsep}{0pt}
        \setlength{\leftmargin 17pt}{\rightmargin 0pt}   
         \setlength{\itemsep}{0pt} \settowidth
        {\labelwidth}{#1.}\sloppy}}{\end{list}}

\def\@citex[#1]#2{\if@filesw\immediate\write\@auxout
        {\string\citation{#2}}\fi
\def\@citea{}\@cite{\@for\@citeb:=#2\do
        {\@citea\def\@citea{,}\@ifundefined
        {b@\@citeb}{{\bf ?}\@warning
        {Citation `\@citeb' on page \thepage \space undefined}}
        {\csname b@\@citeb\endcsname}}}{#1}}
\newif\if@cghi
\def\cite{\@cghitrue\@ifnextchar [{\@tempswatrue
        \@citex}{\@tempswafalse\@citex[]}}
\def\citelow{\@cghifalse\@ifnextchar [{\@tempswatrue
        \@citex}{\@tempswafalse\@citex[]}}
\def\@cite#1#2{{\if@cghi\unskip$\null^{#1}$\else #1\fi\if@tempswa\typeout
        {warning: optional citation argument ignored: `#2'} \fi}}

\def\ps@hep{\addtolength{\headheight}{5pt}
            \addtolength{\topmargin}{-15pt}
            \addtolength{\headsep}{15pt}
    \def\@oddhead{\ifdouble\else\hfil\begin{tabular}{r}
          \texttt{hep-th/9804046 v2}\\ CPT-98/P.3631 \end{tabular}\fi}
          \let\@evenhead\@oddhead
          \def\@oddfoot{\hfil\thepage\hfil}\let\@evenfoot\@oddfoot}
\makeatother

\begin{document}

\thispagestyle{hep}

\begin{center}
{\renewcommand{\thefootnote}{\fnsymbol{footnote}}
{\Large\bfseries $\mathrm{SO(10)}$ unification in noncommutative
geometry revisited}
\\[3ex]
\textsc{Raimar Wulkenhaar}\footnote{supported by the German Academic
Exchange Service (DAAD), grant no.\ D/97/20386. \\
On leave of absence from Institut f\"ur Theoretische Physik,
Universit\"at Leipzig, Augustusplatz 10/11, 04109 Leipzig, Germany.} 
\\[1ex]
{\small{\itshape Centre de Physique Th\'eorique \\
                 CNRS Luminy, Case 907 \\ 
                 13288 Marseille Cedex 9, France }\\
                 e-mail: \texttt{raimar@cpt.univ-mrs.fr}\\[1ex]
                 November 4, 1998} 
\setcounter{footnote}{0}}
\end{center}
 
\renewcommand{\thefootnote}{\alph{footnote}}

\vskip 4ex
 
\begin{abstract}
We investigate the $\mathrm{SO(10)}$ unification model in a
Lie-algebraic formulation of noncommutative geometry. The
$\mathrm{SO(10)}$ symmetry is broken by a $\boldsymbol{45}$-Higgs and
the Majorana mass term for the right neutrinos
($\boldsymbol{126}$-Higgs) to the standard model structure
group. We study the case where the fermion masses are as general as
possible, which leads to two $\boldsymbol{10}$-multiplets, four
$\boldsymbol{120}$-multiplets and two additional
$\boldsymbol{126}$-multiplets of Higgs fields. This Higgs structure
differs considerably from the two Higgs multiplets $\boldsymbol{16}
\otimes \boldsymbol{16}^*$ and $\boldsymbol{16}^c \otimes
\boldsymbol{16}^*$ used by Chamseddine and Fr\"ohlich. We find the usual
tree-level predictions of noncommutative geometry $m_W=\frac{1}{2}
m_t$, $\sin^2 \theta_W =\frac{3}{8}$ and $g_2=g_3$ as well as $m_H
\leq m_t$.
\end{abstract}
 

1998 PACS: 11.15.Ex, 12.10.Dm, 02.20.Sv

\section{Introduction}

Noncommutative geometry (NCG) has a long history in mathematics,
but it were two discoveries by Alain Connes at the end of
the last decade \cite{acf} which made NCG so interesting for physicists: 
\vspace{-\topsep}
\begin{itemize}
\item[--] Dirac operator and smooth functions, both acting on the spinor
Hilbert space, contain all metric information about a spin manifold, 
\vspace{-\parsep}\vspace{-\itemsep}

\item[--] a finite-dimensional version of that setting 
yields some sort of Higgs potential. 
\end{itemize}\vspace{-\topsep}
In the following years, the idea prospered that one should study the
tensor product of ordinary differential geometry and a matrix
geometry. Numerous versions to construct gauge field theories with
spontaneous symmetry breaking (in particular the standard model)
appeared, all more or less inspired by Connes' discovery, but
different in details. These versions developed, competed with each
other, died out -- a perfect example of Darwin's theory of
evolution. Eventually, one version survived \cite{acr}, proposed by
Alain Connes who -- guided by a deep understanding of mathematics~--
completed a set of axioms \cite{acg}. These axioms have their origin
in the classification of classical spin manifolds \cite{dk}. It is
remarkable that the same axioms which are fruitful in the commutative
case of spin manifolds provide the standard model as one of the
simplest noncommutative examples.

Sometimes it happens that branch lines of evolution have attractive 
features, which fascinate mankind long after the death of the
species (think of the dinosaurs). Such a species is grand
unification. The axioms of NCG are compatible with the standard model
but not with grand unification \cite{lmms}. Nevertheless, grand
unification is such an attractive idea that it will not die very soon,
no matter how well the standard model is verified by 
experiment\footnote{Meanwhile the Super-Kamiokande Collaboration
(Phys.\ Rev.\ Lett.\ \textbf{81} (1998) 1562--1567) has reported
evidence for massive neutrinos and, therefore, for a disproof of
the standard model \`a la Connes in its present form. This could be
interpreted as a new boost for grand unification.}.

Technically, the essential progress that NCG brings over the
traditional formulation of gauge theories is that Yang-Mills and Higgs
fields are understood as two complementary parts of one universal
gauge potential.  It is desirable to extend this unifying feature to
grand unified theories (GUTs). The only chance to do so is to allow
for a minor modification of the NCG-axioms so that GUTs are accessible
as well. An inspiration how to modify the axioms comes from
traditional gauge theories. They are formulated in terms of Lie groups
or~-- on infinitesimal level~-- Lie algebras. Thus, replacing the
associative algebra in NCG by a Lie algebra \cite{rw2}, one can expect
a formalism (section \ref{lie}) that is able to produce grand unified
models.

In this paper we show that the $\mathrm{SO(10)}$ GUT \cite{fm,g} can
indeed be formulated with this method. The $\mathrm{SO(10)}$ model has
the exceptional property that all known fermions and the supposed
right-handed neutrino fit into the irreducible 
$\boldsymbol{16}$-representation of $\mathrm{SO(10)}$ (for each
generation). Other GUTs such as $\mathrm{SU(5)}$, $\mathrm{SU(5)}
\times \mathrm{U(1)}$  and $\mathrm{SU(4)} \times \mathrm{SU(2)}_L
\times \mathrm{SU(2)}_R$ arise as intermediate steps of different
$\mathrm{SO(10)}$ symmetry breaking chains. 

The treatment of the $\mathrm{SO(10)}$-model by NCG-methods is not
new. The first approach \cite{cf} by Chamseddine and Fr\"ohlich came
in the early days of noncommutative geometry. Since then, NCG has
undergone the development sketched above that singled out the axioms
incompatible with grand unification. Our construction differs in its
conception and its results from the Chamseddine-Fr\"ohlich
approach. The authors of ref.\ \citelow{cf} start from the
(associative) Clifford algebra of $\mathrm{SO(10)}$. The crucial
difference between the two approaches lies in the Higgs sector. We
denote by $Y=\sum_i Y_i \otimes M^i$ the Yukawa operator. Its part
$Y_i$ transforms the $\boldsymbol{16}$-representation into itself or
into its charge conjugate $\boldsymbol{16}^c$. The mass matrices $M^i$
mix the three fermion generations. Thus, for $Y_i$ we have the two
possibilities $\boldsymbol{16} \otimes \boldsymbol{16}^*$ and
$\boldsymbol{16}^c \otimes \boldsymbol{16}^*$, which are reducible
representations under $\mathrm{SO(10)}$ or $\mathrm{so(10)}$ and
decompose into
\begin{equation}
\boldsymbol{16} \otimes \boldsymbol{16}^* = \boldsymbol{1} \oplus 
\boldsymbol{45} \oplus \boldsymbol{210}~, \qquad 
\boldsymbol{16}^c \otimes \boldsymbol{16}^* = \boldsymbol{10} \oplus 
\boldsymbol{120} \oplus \boldsymbol{126}~. 
\label{decomp}
\end{equation}
The point is now that these two $Y_i$-representations are \emph{not}
reducible under the Clifford algebra $\mathrm{Cliff}(\mathrm{SO(10)})$.
That is why the Higgs multiplets in the Chamseddine-Fr\"ohlich model are 
$\boldsymbol{16} \otimes \boldsymbol{16}^*$ and
$\boldsymbol{16}^c \otimes \boldsymbol{16}$, where only the latter
occurs in the fermionic action (due to chirality). The consequence is
that there is only one generation matrix $M^i$ in the fermionic
action. This leads to the relation 
$m_e : m_\mu : m_\tau = m_d : m_s : m_b = m_u : m_c : m_t$
between the fermion masses,
which is obviously not satisfied. Moreover, the analysis of the Higgs
potential shows that also a Kobayashi-Maskawa matrix is not possible,
but can be obtained by including an additional fermion in the trivial
representation. 

In our version based upon the \emph{Lie algebra} $\mathrm{so(10)}$ we
do have the decomposition \eqref{decomp}, and each irreducible
representation occurring in \eqref{decomp} is tensored by its own
generation matrix. It is therefore no problem to get the fermion
masses we want. For the $\mathrm{SO(10)}$ symmetry breaking we have the
$\boldsymbol{45}$ and $\boldsymbol{210}$-representations at disposal,
which both do not occur in the fermionic action. We employ the 
$\boldsymbol{45}$-representation to break $\mathrm{SO(10)}$ to 
$\mathrm{SU(3)}_C \times \mathrm{SU(2)}_L \times \mathrm{SU(2)}_R \times
\mathrm{U(1)}_{B-L}$ in the first step (at about \cite{m} 
$10^{16} \mathrm{GeV}$). The corresponding (self-adjoint) generation
matrix adds
a freedom of 9 real parameters. The other symmetry breaking
chain $\mathrm{SO(10)} \to \mathrm{SU(4)}_{PS} \times \mathrm{SU(2)}_L
\times \mathrm{SU(2)}_R$ mediated by the $\boldsymbol{210}$ is
possible as well, but we have to make a choice due to the length of
the formulae. In the second step this intermediate symmetry is broken
by the Majorana mass term for the right-handed neutrinos 
($\boldsymbol{126}$-generator) to the standard model symmetry group 
$\mathrm{SU(3)}_C \times \mathrm{SU(2)}_L \times \mathrm{U(1)}_Y$ 
(at about \cite{m} $10^9 \mathrm{GeV}$). 
We then restrict ourselves to the case where the fermion
masses are as general as possible, this implies the Higgs multiplets 
$\boldsymbol{10}$ (twice), $\boldsymbol{120}$ (four times) and
$\boldsymbol{126}$ (twice). The surviving symmetry group is
$\mathrm{SU(3)}_C \times \mathrm{U(1)}_{EM}$. 

There is also a technical difference to mention. The article ref.\ 
\citelow{cf} was written in the pioneering epoch of
noncommutative geometry where auxiliary fields emerged in the
action. They eliminate themselves at the end via their equation of
motion. Our version is based on a differential calculus and we
quotient out the ideal of auxiliary fields before building the
action. We think this method is more transparent but the results are
independent of the way of eliminating these unphysical degrees of
freedom. 

Our paper is organized as follows: We review in section \ref{lie} our
Lie algebraic approach to noncommutative geometry. In section
\ref{sett} we write down our setting of the $\mathrm{SO(10)}$ model, where
we acknowledge a lot of inspiration from ref.\ \citelow{cf}. Section 
\ref{gp} is
devoted to the computation of the gauge potential and the Higgs part
of the field strength. This is the most cumbersome part, because the 
extremely rich Higgs structure leads to a big number of terms in the
field strength. It remains to calculate some traces to get the bosonic
action (section \ref{bose}) and to implement the various symmetries
to get the fermionic action (section \ref{fermi}). We conclude with an
outlook towards a minimal $\mathrm{SO(10)}$ model.

\section{The Lie algebraic formulation of noncommutative geometry}
\label{lie}

The starting point is the Lie algebra $\mathfrak{g}=C^\infty(M) \otimes 
\mathfrak{a}$ acting via a representation $\pi=\mathrm{id} \otimes
\hat{\pi}$ on the Hilbert space $\mathcal{H}=L^2(M,S) \otimes
\mathbb{C}^F$. Here, $C^\infty(M)$ is the algebra of (real-valued)
smooth functions on the (compact Euclidean) spacetime manifold 
$M$, $L^2(M,S)$ is the Hilbert space of square integrable bispinors and 
$\hat{\pi}$ a representation of the semisimple matrix Lie algebra 
$\mathfrak{a}$ on $\mathbb{C}^F$. The treatment of Abelian factors is
possible but more complicated. Moreover, we have the selfadjoint unbounded 
operator $D=\iu \dslash \otimes \one_F + Y$ on $\mathcal{H}$, with 
$Y=\gamma^5 \otimes \hat{Y}$ and $\hat{Y} \in \mathrm{M}_F\mathbb{C}$. 
We also need a $\mathbb{Z}_2$ grading operator $\boldsymbol{\Gamma}$ on 
$\mathcal{H}$ which commutes with $\pi(\mathfrak{g})$ and
anti-commutes with $D$. In many cases there will exist further
discrete symmetries such as  the charge conjugation $J$.

A universal $1$-form $\omega^1 \in \Omega^1$ has the structure 
\[
\omega^1=\sum_{\alpha,z} [f^z_\alpha {\otimes} a^z_\alpha,[\dots 
[f^1_\alpha {\otimes} a^1_\alpha, \mathrm{d} (f^0_\alpha {\otimes}
a^0_\alpha) ]\dots]] ~,
\]
with $f^i_\alpha \in C^\infty(M)$ and $a^i_\alpha \in \mathfrak{a}$. The 
commutators should be read as tensor products. The representation $\pi$ of 
the universal calculus on $\mathcal{H}$ is obtained by taking $\pi$ of 
$f^i_\alpha \otimes a^i_\alpha$ and representing the universal
$\mathrm{d}$ by the derivation $[-\iu D,~.~]$,
\begin{align*}
\rho = \pi(\omega^1) &
\textstyle =\sum_{\alpha,z} [f^z_\alpha {\otimes} 
\hat{\pi}(a^z_\alpha),[\dots [f^1_\alpha {\otimes} \hat{\pi}(a^1_\alpha), 
[-\iu D, f^0_\alpha {\otimes} \hat{\pi}(a^0_\alpha)]]\dots]] 
\\
&\textstyle = \sum_{\alpha,z} f^z_\alpha \dots f^1_\alpha
\dslash(f^0_\alpha) \otimes [\hat{\pi}(a^z_\alpha),
[\dots [\hat{\pi}(a^1_\alpha), \hat{\pi}(a^0_\alpha) ]\dots]] && \to A
\\
&\textstyle + \sum_{\alpha,z} \gamma^5 f^z_\alpha \dots f^1_\alpha
f^0_\alpha  \otimes [\hat{\pi}(a^z_\alpha),[\dots[\hat{\pi}(a^1_\alpha), 
[-\iu \hat{Y},\hat{\pi}(a^0_\alpha)]]\dots]] && \to \pi(\eta)
\end{align*}
The second and third lines are independent for semisimple $\mathfrak{a}$. 
In the second line the commutators clearly yield an element of 
$\hat{\pi}(\mathfrak{a})$ and $f\dslash f'$ a spacetime $1$-form,
together a Yang-Mills multiplet represented on $\mathcal{H}$. We
decompose the finite dimensional part of the Hilbert space into 
$\mathbb{C}^F=\bigoplus_i \boldsymbol{n}_i \otimes \mathbb{C}^N$,
where $\boldsymbol{n}_i$ are irreducible representations of
$\mathfrak{a}$ and $N$ is the number of fermion generations. Then, we
have the decomposition $\hat{Y}=\sum \hat{Y}^r_{ij} \otimes M^{ij}_r$,
where $\hat{Y}^r_{ij} \in \boldsymbol{n}_i \otimes \boldsymbol{n}_j^*$
is (for each $r$) a generator of an irreducible representation and
$M_r^{ij} \in \mathrm{M}_N \mathbb{C}$ a mass matrix. If we now
evaluate the commutators in the third line above, the generators are
expanded to irreducible multiplets, and we obtain  
$\pi(\eta) = \sum \gamma^5 \eta_{ij}^r \otimes M^{ij}_r$, where 
$\eta_{ij}^r \in C^\infty(M) \otimes \boldsymbol{n}_{ij}$ are
function-valued irreducible representations, i.e.\ Higgs multiplets.

The universal differential of $\omega^1$ is defined as 
\[
\mathrm{d}\omega^1= \sum_{\alpha,z} \sum_{y=1}^z 
[f^z_\alpha {\otimes} a^z_\alpha,[\dots 
[\mathrm{d}(f^y_\alpha {\otimes} a^y_\alpha),[\dots 
[f^1_\alpha {\otimes} a^1_\alpha, \mathrm{d} (f^0_\alpha {\otimes} 
a^0_\alpha) ]\dots]]\dots]] ~.
\]
Its representation on $\mathcal{H}$ reads after elementary calculation
\begin{align}
\pi(\mathrm{d}\omega^1) 
&= \!\! \sum_{\alpha,z} \sum_{y=1}^z 
[f^z_\alpha {\otimes} \hat{\pi}(a^z_\alpha),[\dots 
\{ [-\iu D, f^y_\alpha {\otimes} \hat{\pi}(a^y_\alpha)],[\dots 
[-\iu D, f^0_\alpha {\otimes} \hat{\pi}(a^0_\alpha)]\dots]\} \dots ]] 
\notag \\
&=\{-\iu D,\rho\} + \textstyle 
\sum_{\alpha,z} [f^z_\alpha {\otimes} 
\hat{\pi}(a^z_\alpha),[\dots [f^1_\alpha {\otimes} \hat{\pi}(a^1_\alpha), 
[D^2, f^0_\alpha {\otimes} \hat{\pi}(a^0_\alpha)]]\dots]] 
\notag \\
&\equiv \{-\iu D,\pi(\omega^1)\} + \sigma(\omega^1) \label{d1}
\\
&= \{\dslash,A\} \textstyle 
+ \sum_{\alpha,z} [f^z_\alpha {\otimes} 
\hat{\pi}(a^z_\alpha),[\dots [f^1_\alpha {\otimes} \hat{\pi}(a^1_\alpha), 
[-\dslash^2 \otimes \one_F, f^0_\alpha {\otimes} \hat{\pi}(a^0_\alpha)]]
\dots]] \notag
\\
& + \{\dslash,\pi(\eta)\} + \{-\iu Y,A\}
+ \{-\iu Y,\pi(\eta)\} + \hat{\sigma}(\eta)~, \notag
\end{align}
with 
\[
\textstyle 
\hat{\sigma}(\eta) = \sum_{\alpha,z} f^z_\alpha \dots f^1_\alpha 
f^0_\alpha \otimes [\hat{\pi}(a^z_\alpha),[\dots [\hat{\pi}(a^1_\alpha),
[\hat{Y}^2,\hat{\pi}(a^0_\alpha)]]\dots]] ~.
\]
After a lengthy calculation \cite{rw2} one finds 
\begin{align*}
\textstyle \{\dslash,A\} + \sum_{\alpha,z} [f^z_\alpha {\otimes} &
\hat{\pi}(a^z_\alpha),[\dots [f^1_\alpha {\otimes} \hat{\pi}(a^1_\alpha), 
[-\dslash^2 \otimes \one_F, f^0_\alpha {\otimes} \hat{\pi}(a^0_\alpha)]]
\dots]] \\ &= \mathbf{d} A + C^\infty(M) \otimes 
\{\hat{\pi}(\mathfrak{a}),\hat{\pi}(\mathfrak{a})\}~,
\end{align*}
where $\mathbf{d}$ is the exterior differential (which anti-commutes
with $\gamma^5$) and the
$\{\hat{\pi}(\mathfrak{a}),\hat{\pi}(\mathfrak{a})\}$ part is
independent of $A$ and $\pi(\eta)$. Moreover, we have
$\{\dslash,\pi(\eta)\} = \mathbf{d} \pi(\eta)$.  We now decompose
$\hat{Y}^2$ into generators of irreducible representations,
\[
\hat{Y}^2=\hat{Y}^2_{\|} + \hat{Y}^2_\perp + \hat{\pi}(\boldsymbol{1})~.
\]
Here, $\hat{\pi}(\boldsymbol{1})$ contains trivial representations which 
commute with $\hat{\pi}(\mathfrak{a})$. Those generators which also 
occur in $\hat{Y}$, denoted as $\hat{Y}^2_{\|}$, generate obviously the 
corresponding representations which already occur in $\pi(\eta)$. The 
other (non-trivial) generators, denoted $\hat{Y}^2_\perp$, generate 
representations independent of $\pi(\eta)$ and $A$. 

It is now crucial to note that the same $\rho$ can be written in many ways 
as $\pi(\omega^1)$ so that the definition of a differential 
``$d\rho=\pi(\mathrm{d}\omega^1)$'' is ambiguous. The usual way out is to 
consider equivalence classes modulo the ideal 
$\mathcal{J}^2=\pi(\mathrm{d} (\ker \pi \cap \Omega^1))$. We have just 
shown that if $\pi(\omega^1)=0$ then there remain only the 
$\{\hat{\pi}(\mathfrak{a}),\hat{\pi}(\mathfrak{a})\}$ part and the 
representations generated by $\hat{Y}^2_\perp$, which gives
\begin{equation}
\textstyle \mathcal{J}^2 = C^\infty(M) \otimes \sum 
(\{\hat{\pi}(\mathfrak{a}), \hat{\pi}(\mathfrak{a})\} + 
[\hat{\pi}(\mathfrak{a}),[\dots [\hat{\pi}(\mathfrak{a}), 
\hat{Y}^2_\perp]\dots]] )~.
\end{equation}
The final formula for the differential of $\rho=A+\pi(\eta)$ is therefore
\begin{equation}
d \rho = \mathbf{d} \rho + \{-\iu Y,\rho\} + \hat{\sigma}(\eta) 
\mod \mathcal{J}^2~.
\label{diff}
\end{equation}
In the same way one represents the space $\Omega^n$ of universal forms of 
degree $n$ on $\mathcal{H}$ and determines the corresponding ideal 
$\mathcal{J}^n = \pi(\mathrm{d} (\ker \pi \cap \Omega^{n-1}))$. The
generalization of \eqref{d1} is 
\begin{equation}
d(\pi(\omega^n)+ \mathcal{J}^n)= [\![-\iu D,\pi(\omega^n)]\!] 
+ \sigma(\omega^n) + \mathcal{J}^{n+1}~,\qquad \omega^n \in \Omega^n~,
\label{ds}
\end{equation}
where $[\![~.~,~.~]\!]$ is the graded commutator, i.e.\ the
anti-commutator if both entries are odd under $\mathbb{Z}_2$ and the
commutator else. This yields the graded differential Lie algebra 
$\Omega_D = \pi(\Omega)/\mathcal{J}$. 

We propose to define the connection $\nabla$ as a generalization of the
differential $d$ and the covariant derivative $\mathcal{D}$ as a
generalization of the operator $D$. This means that $\mathcal{D}$ is
a linear unbounded selfadjoint operator on $\mathcal{H}$ and odd
under $\mathbb{Z}_2$, and $\nabla: \Omega^n_D \to \Omega_D^{n+1}$ is
linear. Both $\mathcal{D}$ and $\nabla$ are related via the same
formula \eqref{ds}:
\[
\nabla(\pi(\omega^n) + \mathcal{J}^n) 
= [\![-\iu \mathcal{D},\pi(\omega^n)]\!]  
+ \sigma(\omega^n) + \mathcal{J}^{n+1}~,
\]
for $\omega^n \in \Omega^n$ and any degree $n$. The general solution
is  
\[
\mathcal{D} = D + \iu \rho~,\quad \nabla= d + [\![\rho,~.~]\!]~,\quad 
[\![\rho, \pi(\Omega^n)]\!] \subset \pi(\Omega^{n+1})~, \quad 
[\![\rho, \mathcal{J}^n]\!] \subset \mathcal{J}^{n+1}~.
\]
One obvious solution is $\rho = A + \pi(\eta) \in \pi(\Omega^1)$. But 
there are further solutions possible, depending on the setting. These
additional solutions allow us to formulate gauge theories with
$\mathrm{u(1)}$ factors such as the standard model. Demanding that 
$\rho$ commutes with functions, we have the decomposition
\[
\rho' \in \Lambda^1 \otimes \mathbf{r}^0 + \Lambda^0 \gamma^5 \otimes 
\mathbf{r}^1
\]
of the additional solutions, where the matrices $\mathbf{r}^i 
\in \mathrm{M}_F \mathbb{C}$ commute with $\hat{\pi}(\mathfrak{a})$.
The essential step is to check $\{\rho',\pi(\Omega^1)\} 
\subset \pi(\Omega^2)$, which yields several conditions for $\mathbf{r}^0$ 
and $\mathbf{r}^1$. Finally, one has to verify the compatibility 
with $\mathcal{J}$. 

The curvature is now $\nabla^2=[\mathcal{F},~.~]$, where one finds the 
usual formula
\[
\mathcal{F} = d\rho + \tfrac{1}{2} \{ \rho,\rho\}
\]
for the field strength. The differential and (anti)commutator are
defined via  the (graded) Leibniz rule and Jacobi identity; for it one
has to enlarge the  ideal $\mathcal{J}$ by the graded centralizer
$\mathcal{C}$ of $\pi(\Omega)$. Then, the general formula \eqref{diff}
continues to work so that for  $\rho=A+\pi(\eta)$ one has 
\begin{align}
\mathcal{F} &= (\mathbf{d}A + A^2|_{\Lambda^2}) 
+ (\mathbf{d} \pi(\eta) + \{A,(\pi(\eta)-\iu Y)\}) \notag \\ & 
+ \big((\pi(\eta))^2+\{-\iu Y,\pi(\eta)\} +\hat{\sigma}(\eta) 
\mod \mathcal{J}^2 + \mathcal{C}^2 \big)~. \label{F}
\end{align}
Here, $A^2|_{\Lambda^2}$ is the restriction of $A^2$ to the 
spacetime 2-form part. The bosonic action is defined via the Dixmier trace 
and can be rewritten as 
\begin{align}
S_B &= \tfrac{1}{g^2\,F} \int_M dx \; \mathrm{tr}(\mathcal{F}_\perp)^2  
= \int_M dx \; (\mathcal{L}_2 + \mathcal{L}_1 + \mathcal{L}_0) 
\label{SB} \\
&= \tfrac{1}{g^2\,F} \int_M dx \begin{array}[t]{l} \Big( 
\mathrm{tr}((\mathbf{d}A + A^2|_{\Lambda^2})^2) 
+ \mathrm{tr}((\mathbf{d} \pi(\eta) + \{A,(\pi(\eta)-\iu Y)\})^2) \notag \\
+ \mathrm{tr}(\big((\pi(\eta))^2+\{-\iu Y,\pi(\eta)\} +\hat{\sigma}(\eta)
\big)_\perp^2) \Big)~. \end{array} \notag
\end{align}
Here, $g$ is a coupling constant and $F$ the dimension of the matrix part. 
The trace includes the trace over gamma matrices and $\mathcal{F}_\perp$ 
is the component of $\mathcal{F}$ orthogonal to $\mathcal{J}^2$. The bosonic 
action consists of three parts, the Yang-Mills Lagrangian
$\mathcal{L}_2$, the covariant derivative $\mathcal{L}_1$ of the Higgs
fields and the Higgs  potential $\mathcal{L}_0$. The fermionic action is 
\begin{equation}
S_F = \tfrac{1}{2^s} \int_M dx\; \boldsymbol{\psi}^* \mathcal{D} 
\boldsymbol{\psi} 
= \tfrac{1}{2^s} \int_M dx\; \iu \boldsymbol{\psi}^* (\dslash \otimes \one_F 
+ A + \pi(\eta) - \iu Y) \boldsymbol{\psi}~, 
\label{SF}
\end{equation}
where $s$ is the number of discrete symmetries of the setting, and 
$\boldsymbol{\psi} \in \mathcal{H}$ is invariant under these 
symmetries (possibly only after passing to Minkowski space).
The fermionic action contains the minimal coupling to the Yang-Mills fields 
$A$ and the Yukawa coupling to the Higgs fields $\pi(\eta)$.

Let us study the part $\theta=\big((\pi(\eta))^2+\{-\iu Y,\pi(\eta)\} 
+\hat{\sigma}(\eta)\big)_\perp$ of the field strength, whose square gives 
the Higgs potential. Since the covariant derivative $\mathcal{D}$ can be 
written as $\mathcal{D} = \iu \dslash \otimes \one_F + \iu A 
+ \iu (\pi(\eta)-\iu Y)$, it is natural to consider 
$\pi(\tilde{\eta})=(\pi(\eta)-\iu Y)$ as the analogue of a classical Higgs 
multiplet. This is because $\pi(\tilde{\eta})$ 
transforms as $\pi(\tilde{\eta}) \mapsto u\pi(\tilde{\eta}) u^*$ under 
gauge transformations $\mathcal{D} \mapsto u\mathcal{D} u^*$, 
with $u \in C^\infty(M) \otimes \exp(\mathfrak{a}+\mathbf{r}^0)$. In
terms of $\tilde{\eta}$ we have
\begin{align}
\theta &=(\pi(\tilde{\eta})^2 + \hat{\sigma}(\eta) + Y^2)_\perp = 
(\pi(\tilde{\eta})^2 + \hat{\sigma}(\eta) + \pi(\boldsymbol{1}) 
+ Y^2_\perp + Y^2_{\|})_\perp \notag \\
&= (\pi(\tilde{\eta})^2 + \hat{\sigma}(\tilde{\eta}) + \pi(\boldsymbol{1}) 
)_\perp ~, \label{th}
\end{align}
because $Y^2_\perp$ has by definition no component orthogonal to 
$\mathcal{J}^2$ and 
\begin{align*}
\mbox{if} &&
\pi(\tilde{\eta}) &=-\iu Y 
\textstyle + \sum_{\alpha,z} [\pi(a^z_\alpha),[\dots [\pi(a^1_\alpha),
[-\iu Y,\pi(a^0_\alpha)]]\dots]] 
\\
\mbox{then} &&
\hat{\sigma}(\tilde{\eta}) &=Y^2_{\|}
\textstyle + \sum_{\alpha,z} [\pi(a^z_\alpha),[\dots [\pi(a^1_\alpha),
[Y^2_{\|},\pi(a^0_\alpha)]]\dots]] ~,
\end{align*}
with $a^i_\alpha \in \mathfrak{g}$. Thus, 
$\theta$ can be expressed completely in terms of $\tilde{\eta}$ so that 
gauge invariance of the Higgs potential $V=\mathrm{tr}(\theta^2)$ is obvious. 
The point is now that we know a priori the Higgs vacuum: it is 
$\langle \pi(\tilde{\eta}) \rangle_0= -\iu Y$. At this configuration we 
have $\theta=0$ and 
$V=0$. On the other hand $V$ is non-negative so that $-\iu Y$ is a global 
and local minimum. In the vicinity of $-\iu Y$ we have 
$V=\mathrm{tr}((-\iu (Y \pi(\eta)+\pi(\eta) Y) + \hat{\sigma}(\eta)
)_\perp^2)$, i.e.\ something bilinear in the physical Higgs fields. 
The coefficients are the Higgs masses, after diagonalization and rescaling. 
There are however massless modes, the Goldstone bosons. They are of the
form $\pi(\eta)=[\pi(a),-\iu Y]$ with $a \in \mathfrak{g}$. 
In this case we have 
\begin{align*}
(-\iu (Y \pi(\eta)+\pi(\eta) Y) + \hat{\sigma}(\eta))_\perp 
&= (-[\pi(a),Y^2] + [\pi(a),Y^2_{\|}])_\perp \\
&= (-[\pi(a),Y^2_\perp])_\perp = 0~.
\end{align*}
The masses of the Yang-Mills fields come from the part 
$\mathrm{tr}(\{A,-\iu Y\})^2 = \mathrm{tr}([\pi(A_\mu),Y][\pi(A^\mu),Y])$ 
of the Lagrangian $\mathcal{L}_1$. This is a form of the Goldstone-Higgs 
theorem: The massive Yang-Mills fields are those which do not commute with 
$Y$ and to each of them there corresponds a massless Goldstone boson. The 
Higgs mechanism consists in removing the Goldstone bosons by those gauge 
transformations which do not commute with $Y$, and which are fixed in this 
way. The remaining unconstrained gauge degrees of freedom are those which 
commute with $Y$, and to each of them there corresponds a massless
Yang-Mills field.

\section{The $\mathrm{so(10)}$ setting}
\label{sett}

Let $\Gamma_I \in \mathrm{M}_{32}\mathbb{C}$, $I=0,\dots, 9$, be
$\mathrm{so(10)}$ gamma matrices represented in terms of tensor
products of five sets of Pauli matrices \cite{cf}
\begin{align}
\Gamma_i &=\kappa_1 \rho_3\eta_i ~, &
\sigma_i &\to \one_2 \otimes \one_2 \otimes \one_2 \otimes \one_2 \otimes 
              \sigma_i ~, \notag \\
\Gamma_{i+3} &= \kappa_1 \rho_1 \sigma_i ~, &
\tau_i & \to \one_2 \otimes \one_2 \otimes \one_2 \otimes \tau_i \otimes 
               \one_2 ~, \notag \\
\Gamma_{i+6} &= \kappa_1 \rho_2 \tau_i ~, &
\eta_i & \to \one_2 \otimes \one_2 \otimes \eta_i \otimes \one_2 \otimes 
               \one_2 ~, \\
\Gamma_0 &= \kappa_2 ~, &
\rho_i & \to \one_2 \otimes \rho_i \otimes \one_2 \otimes \one_2 \otimes 
               \one_2 ~, \notag \\
\Gamma_{11} &= \iu \Gamma_0 \Gamma_1 \cdots \Gamma_9 = \kappa_3 ~, &
\kappa_i & \to \kappa_i \otimes \one_2 \otimes \one_2 \otimes \one_2 \otimes 
               \one_2 ~, \notag
\end{align}
where $i=1,2,3$. The tensor products are interpreted such that $\sigma_i$ is 
$2 \times 2$ and $\kappa_i$ is $32 \times 32$.

The matrix Lie algebra is $\mathfrak{a}=\mathrm{so(10)}$ represented
as $\mathrm{so(10)} \ni a = a^{IJ} \Gamma_{IJ}$, with
$a^{IJ}=-a^{JI}\in \mathbb{R}$, and where
$\Gamma_{I_1 I_2 \dots I_n}= (1/n!) \Gamma_{\lbrack I_1} \Gamma_{I_2}
\cdots \Gamma_{I_n\rbrack}$
is the completely anti-symmetrized product of $\Gamma$ matrices. 
Summation over equal $\mathrm{so(10)}$ indices $I,J, \dots $ from 0 to 9 
and over equal spacetime indices $\kappa,\lambda, \dots$ from $0$ to $3$ is 
understood. We introduce the projection operators 
\[
P_\pm=\tfrac{1}{2} (1\pm \Gamma_{11}) \otimes \one_3 ~,\quad 
\mathcal{P}=\mathrm{diag}(\one_4 \otimes P_+ \,,\, \one_4 \otimes P_+ \,,\, 
\one_4 \otimes P_- \,,\, \one_4 \otimes P_- )~,
\]
and the $\mathrm{so}(10)$ conjugation matrix
\begin{align*}
B &=-\Gamma_1\Gamma_3\Gamma_4\Gamma_6\Gamma_8 \otimes \one_3
=\bar{B}=B^T=B^\dagger~, & B^2 &=\one_{96}~,
\\ 
B (\Gamma_I \otimes m) B &= \Gamma_I^T \otimes m~~
\forall m \in \mathrm{M}_3\mathbb{C}~,  & B P_\pm B &=P_\mp~,
\end{align*}
and define the $\mathbb{Z}_2$ grading operator 
\[
\boldsymbol{\Gamma}=\mathrm{diag}(-\gamma^5 \otimes P_+ \,,\, 
\gamma^5 \otimes P_+ \,,\, \gamma^5 \otimes P_- \,,\, 
-\gamma^5 \otimes P_- )~.
\]
Then, the Hilbert space is 
\begin{equation}
\mathcal{H}=\mathcal{P} (L^2(M,S) \otimes \mathbb{C}^{32}\otimes
\mathbb{C}^3 \otimes \mathbb{C}^4 ) 
\cong L^2(M,S) \otimes \mathbb{C}^{192} ~.
\end{equation}
The representation $\pi$ of $\mathfrak{g}=C^\infty(M) \otimes \mathfrak{a}
\ni f \otimes a$ on $\mathcal{H}$ is defined by
\begin{equation}
\label{A}
{\arraycolsep 0pt
\pi(f \otimes a)=f\one_4 \otimes \left( \begin{array}{cccc} 
P_+ (a \otimes \one_3)P_+ & 0 & 0 & 0 \\
0 & P_+ (a \otimes \one_3) P_+  & 0 & 0 \\
0 & 0 & P_- (a \otimes \one_3)P_- & 0 \\
0 & 0 & 0 & P_- (a \otimes \one_3) P_- \end{array} \right).
}
\end{equation}
Note that $P_\pm$ commutes with $a \otimes \one_3$ and that
$\boldsymbol{\Gamma}$ commutes with $\pi(a)$. The selfadjoint Yukawa
operator anti-commuting with $\boldsymbol{\Gamma}$ is 
\begin{equation}
Y=\left( \!\!\! \begin{array}{cccc} 0 & 
\gamma^5 \otimes P_+ \mathcal{M} P_+ & 
\gamma^5 \otimes P_+ \mathcal{N} P_-& 0 \\ 
\gamma^5{}^* \otimes P_+ \mathcal{M} P_+ & 0 & 0 & 
\gamma^5{}^* \otimes P_+ \mathcal{N} P_- \\
\gamma^5{}^* \otimes P_- \mathcal{N}^\dagger P_+ & 0 & 0 & 
\gamma^5{}^* \otimes P_- B \overline{\mathcal{M}} B P_- \\ 
0 & \gamma^5 \otimes P_- \mathcal{N}^\dagger P_+ & 
\gamma^5 \otimes P_- B \overline{\mathcal{M}} B P_- & 0 
\end{array} \!\!\! \right) .
\end{equation}
We distinguish explicitly $\gamma^5$ and $\gamma^5{}^*$, which are equal in 
Euclidean space, in Minkowski space however we have $\gamma^5=-\gamma^5{}^*$. 
This allows for a parallel development of our model in both Euclidean and 
Minkowski space. The matrices $\mathcal{M}$ and $\mathcal{N}$ are
obtained by tensoring the generators given in ref.\ \citelow{cf} by
\emph{independent} generation matrices: 
\begin{align}
\mathcal{M} &= - \iu (\Gamma_{45} + \Gamma_{78} +
\Gamma_{69}) \otimes M_1~, \notag \\
\mathcal{N} &= \iu \Gamma_0 \otimes M_s + \Gamma_3 \otimes M_p +
\Gamma_{120} \otimes M_a' - \iu \Gamma_{123} \otimes M_a \\ & 
+ (\Gamma_{450} + \Gamma_{780} + \Gamma_{690}) \otimes M_b' 
- \iu (\Gamma_{453} + \Gamma_{783} + \Gamma_{693}) \otimes M_b \notag \\ &
- \iu (\Gamma_{01245} + \Gamma_{01278} + \Gamma_{01269}) \otimes M_c 
- (\Gamma_{31245} + \Gamma_{31278} + \Gamma_{31269}) \otimes M_f
\notag \\ &
- \tfrac{1}{8} \iu (\Gamma_1 {-} \iu \Gamma_2) \Gamma_3 
(\Gamma_4 {-} \iu \Gamma_5)(\Gamma_6 {-} \iu \Gamma_9)
(\Gamma_7 {-} \iu \Gamma_8) \otimes M_2 ~.\notag 
\end{align}
Here, $M_s,M_p,M_c,M_f,M_2$ are symmetric and $M_a,M_a',M_b,M_b'$
anti-symmetric \mbox{$3 \times 3$}-matrices and $M_1=M_1^\dagger$.
That implies $B\mathcal{N}B=\mathcal{N}^T$ and 
$\mathcal{M}=\mathcal{M}^\dagger$. We stress that here lies the
essential difference between our Lie formalism and the algebraic 
version \cite{cf}: There, the matrices $M_s,M_p,M_c,M_f,M_2$ are all
proportional to each other, the same is true for the matrices 
$M_a,M_a',M_b,M_b'$. This is dictated by the fact that
$\boldsymbol{16}^c \otimes \boldsymbol{16}^*$ is irreducible under the
Clifford algebra of $\mathrm{SO(10)}$.

The above setting is chosen in such a way that it has two symmetries
$J$ and $\mathcal{S}$. First, the charge conjugation is given by 
\[
{\arraycolsep 1pt
J=\mathcal{P}\left( \begin{array}{cccc} 0 & 0 & C {\otimes} B & 0 \\
0 & 0 & 0 & C {\otimes} B \\
C {\otimes} B & 0 & 0 & 0 \\
0 & C {\otimes} B & 0 & 0 \end{array} \right) \mathcal{P} \circ c.c~,
}\]
where $C$ is the spacetime conjugation matrix, 
$C\overline{\gamma^\mu} C=\gamma^\mu$, and $c.c$ stands for complex 
conjugation. We use the following convention for Euclidean gamma matrices
\begin{align*}
\gamma^0 &= \left(\!\!\! \begin{array}{cc} 0 & \one_2 \\ \one_2 & 0 
\end{array} \!\!\! \right), &
\gamma^a &= \left( \!\!\! \begin{array}{cc} 0 & \iu \sigma^a \\ 
-\iu \sigma^a & 0 \end{array} \!\!\! \right), &
\gamma^5 &= \gamma^0 \gamma^1 \gamma^2 \gamma^2 
= \left( \!\!\! \begin{array}{cc} 
\one_2 & 0 \\ 0 & -\one_2 \end{array} \!\!\! \right), & 
C=\gamma^0 \gamma^2\,.
\end{align*}
Our Minkowskian gamma matrices are 
\begin{align*}
\gamma^0 &= \left( \!\!\! \begin{array}{cc} 0 & \one_2 \\ \one_2 & 0 
\end{array} \!\!\! \right) , &
\gamma^a &= \left( \!\!\! \begin{array}{cc} 0 & - \sigma^a \\ \sigma^a & 0 
\end{array} \!\!\! \right) , &
\gamma^5 &= \iu \gamma^0 \gamma^1 \gamma^2 \gamma^3 
= \left( \!\!\! \begin{array}{cc} 
\one_2 & 0 \\ 0 & -\one_2 \end{array} \!\!\! \right), & 
C=\gamma^2\,.
\end{align*}
Observe that $J^2=\mathcal{P}$ in Minkowski space but $J^2=-\mathcal{P}$ 
in Euclidean space. Second, we have an exchange symmetry
\[ 
{\arraycolsep 1pt
\mathcal{S}_E= \mathcal{P} \left( \begin{array}{cccc} 
0 & \one_{384} & 0 & 0 \\
\one_{384} & 0 & 0 & 0 \\
0 & 0 & 0 & \one_{384} \\
0 & 0 & \one_{384} & 0 \end{array} \right) \mathcal{P} ,\quad 
\mathcal{S}_M= \mathcal{P} \left( \begin{array}{cccc} 
0 & \iu \one_{384} & 0 & 0 \\
-\iu \one_{384} & 0 & 0 & 0 \\
0 & 0 & 0 & -\iu \one_{384} \\
0 & 0 & \iu \one_{384} & 0 \end{array} \right) \mathcal{P} ,
}
\]
where $\mathcal{S}_E$ is realized in Euclidean space and 
$\mathcal{S}_M$ in Minkowski space. This yields in both Euclidean 
and Minkowski spaces 
\[
[J,D]=[J,\pi(a)]=[\mathcal{S},D]=[\mathcal{S},\pi(a)]=[J,\mathcal{S}]=0~,
\]
where $D=\mathcal{P} (\iu \dslash \otimes \one_{32} \otimes \one_3 
\otimes \one_4) \mathcal{P} + Y$. 

\section{The gauge potential and its field strength}
\label{gp}

The gauge potential $\rho \in \pi(\Omega^1)$ is composed 
of two parts, of a $\mathfrak{a}$-valued spacetime 1-form 
$A=\gamma^\mu \pi(A_\mu)$ and (up to $\gamma^5$) an 
$\mathfrak{a}$-representation-valued spacetime 0-form $\pi(\eta)$, 
$\rho=A+ \pi(\eta)$. The second part has the general structure
\[
\pi(\eta)=\sum_{\alpha,z}[\pi(a^z_\alpha),[ \dots, [\pi(a^1_\alpha), 
[-\iu Y,\pi(a^0_\alpha)]]\dots]]~,
\]
where $a^i_\alpha \in \mathfrak{g}$. Products of $\Gamma$ matrices are
generators of irreducible representations, but as 
some of them occur more than once in $Y$, we must check that they are linear
independent. For instance,
\[
-\tfrac{1}{4} \mathrm{ad}_{\Gamma_{01}} \circ
\mathrm{ad}_{\Gamma_{01}}(\Gamma_0) \equiv - \tfrac{1}{4}
[\Gamma_{01},[\Gamma_{01},\Gamma_0]]=\Gamma_0~,\quad 
- \tfrac{1}{4} \mathrm{ad}_{\Gamma_{01}} \circ
\mathrm{ad}_{\Gamma_{01}}(\Gamma_3)=0~, 
\]
which establishes the independence of the two 10-dimensional
representations generated by $\Gamma_0 \otimes M_s$ and $\Gamma_3
\otimes M_p$. Next, application of $- \tfrac{1}{4}
\mathrm{ad}_{\Gamma_{01}} \circ \mathrm{ad}_{\Gamma_{01}}$ to the four
120-dimensional representations
generated by $\Gamma_{IJK}$ establishes the independence of 
$\Gamma_{123}$ and $(\Gamma_{450}+\Gamma_{780}+\Gamma_{690})$ from 
$\Gamma_{120}$ and
$(\Gamma_{453}+\Gamma_{783}+\Gamma_{693})$. Application of 
$\tfrac{1}{8} \mathrm{ad}_{\Gamma_{16}} \circ
\mathrm{ad}_{\Gamma_{64}} \circ \mathrm{ad}_{\Gamma_{41}}$ leads to
independence of all these four 120-dimensional
representations. Finally, application of  
$-\tfrac{1}{4} \mathrm{ad}_{\Gamma_{69}} \circ
\mathrm{ad}_{\Gamma_{69}}$ and $-\tfrac{1}{4}
\mathrm{ad}_{\Gamma_{01}} \circ \mathrm{ad}_{\Gamma_{01}}$  
to the three 126-dimensional representations generated by
$\Gamma_{IJKLM}$ shows that they are independent. In conclusion, and
using the identity $B(\overline{a \otimes \one_3})B=a\otimes \one_3$,
the general form of $\pi(\eta)$ is 
\begin{align}
\pi(\eta) &= -\iu \mathcal{P} \left( \!\! \begin{array}{cccc} 
0 & \gamma^5 \: \eta_\mathcal{M} & 
\gamma^5 \: \eta_\mathcal{N} & 0 \\
\gamma^5{}^* \: \eta_\mathcal{M} & 0 & 0 & 
\gamma^5{}^* \: \eta_\mathcal{N} \\
\gamma^5{}^* \: \eta_\mathcal{N}^\dagger & 0 & 0 & 
\gamma^5{}^* \: B \overline{\eta_\mathcal{M}} B  \\
0 & \gamma^5 \: \eta_\mathcal{N}^\dagger & 
\gamma^5 \: B \overline{\eta_\mathcal{M}} B & 0 
\end{array} \right) \mathcal{P}\,,
\\
\eta_\mathcal{M} &= - \iu \Theta \otimes M_1 \notag \\
\eta_\mathcal{N} &= \iu \Upsilon_1 \otimes M_s + \Upsilon_2 \otimes M_p 
+ \Phi_1 \otimes M_a' - \iu \Phi_2 \otimes M_a + \Phi_3 \otimes M_b' 
- \iu \Phi_4 \otimes M_b \notag \\ & 
- \iu \Psi_1 \otimes M_c - \Psi_2 \otimes M_f - \iu \Psi_3 \otimes M_2 ~, 
\notag
\end{align}
where $\Theta \in C^\infty(M) \otimes \boldsymbol{45}$, 
$\Upsilon_i \in C^\infty(M) \otimes \boldsymbol{10}$ and 
$\Phi_i \in C^\infty(M) \otimes \boldsymbol{120}$, all 
of them being real representations, and 
$\Psi_i \in C^\infty(M) \otimes \boldsymbol{126}$ 
(complex representation). 

The next step is to compute the ideal $\mathcal{J}^2$ and
the part $\hat{\sigma}(\eta)$ of the curvature. Both are related to
$Y^2$, decomposed into irreducible representations. Those
which occur in both $Y$ and $Y^2$ contribute to
$\hat{\sigma}(\eta)$, the others give rise to the ideal. Using
$\gamma^5 \gamma^5{}^*=\epsilon \one_4$, with $\epsilon=1$ in Euclidean 
space and $\epsilon=-1$ in Minkowski space, we have 
\begin{align}
Y^2 &=\epsilon \one_4 \otimes \left( \begin{array}{cccc} 
P_+ \mathcal{Y}_{(1)} P_+ & 0 & 0 & \mathcal{Y}_{(2)} P_- \\
0 & P_+ \mathcal{Y}_{(1)} P_+ & P_+ \mathcal{Y}_{(2)} P_- & 0 \\
0 & P_- \mathcal{Y}_{(2)}^\dagger P_+ & 
P_- B \overline{\mathcal{Y}_{(1)}} B P_- & 0 \\
P_- \mathcal{Y}_{(2)}^\dagger P_+ & 0 & 0 & 
P_- B \overline{\mathcal{Y}_{(1)}} B P_- \end{array} \right) ,
\\
\mathcal{Y}_1 &=\mathcal{M}^2 + \mathcal{N} \mathcal{N}^\dagger~,\qquad 
\mathcal{Y}_2 =\mathcal{M}\mathcal{N} + \mathcal{N}
B \overline{\mathcal{M}} B~. \notag
\end{align}
In detail, we find
\begin{align*}
\mathcal{Y}_{(1)} &=\one_{32}\otimes (3 M_1 M_1^\dagger + M_s M_s^\dagger
+ M_p M_p^\dagger + M_a' M_a'{}^\dagger + M_a M_a^\dagger \\ &
{}\quad + 3 M_b' M_b'{}^\dagger + 3 M_b M_b^\dagger + 3 M_c M_c^\dagger 
+ 3 M_f M_f^\dagger + \tfrac{1}{4} M_2 M_2^\dagger)
\\ &
- \iu (\Gamma_{45} + \Gamma_{78} + \Gamma_{69}) \otimes 
(M_s M_b'{}^\dagger\!
+ M_b' M_s^\dagger \!+ M_p M_b^\dagger \!+ M_b M_p^\dagger\! 
+ M_a' M_c^\dagger \!+ M_c M_a'{}^\dagger \\ & {}\quad 
+ M_a M_f^\dagger + M_f M_a^\dagger + 2 M_b' M_f^\dagger 
+ 2 M_f M_b'{}^\dagger 
+ 2 M_b M_c^\dagger + 2 M_c M_b^\dagger - \tfrac{1}{4} M_2 M_2^\dagger)
\\ &
+ \iu \Gamma_{03} \otimes \mathcal{Z}_1 
- \iu \Gamma_{12} \otimes \mathcal{Z}_2 
\\ &
- (\Gamma_{4578} + \Gamma_{4569} + \Gamma_{7869}) \otimes \mathcal{Z}_3
+ (\Gamma_{0345} + \Gamma_{0378} + \Gamma_{0369}) \otimes \mathcal{Z}_4
\\ &
- (\Gamma_{1245} + \Gamma_{1278} + \Gamma_{1269}) \otimes
\mathcal{Z}_5
+ \Gamma_{0123} \otimes \mathcal{Z}_6
\\ &
+ \tfrac{1}{8} (\Gamma_{2567} - \Gamma_{1467} + \Gamma_{1568} 
+ \Gamma_{2468} - \Gamma_{1489} - \Gamma_{2479} + \Gamma_{2589} -
\Gamma_{1579}) \otimes \mathcal{Z}_7
\\ & 
+ \tfrac{1}{8} (\Gamma_{1567} + \Gamma_{2467} - \Gamma_{2568} +
\Gamma_{1468} + \Gamma_{2489} - \Gamma_{1479} + \Gamma_{1589} +
\Gamma_{2579}) \otimes \mathcal{Z}_8~.
\\[1ex]
\mathcal{Y}_{(2)} &= 
\iu \Gamma_{0} \otimes (3 M_1 M_b' - 3 M_b' \overline{M}_1) 
+ \Gamma_{3} \otimes (3 M_1 M_b - 3 M_b \overline{M}_1) 
\\ & 
+ \Gamma_{120} \otimes (3 M_1 M_c - 3 M_c \overline{M}_1) 
- \iu \Gamma_{123} \otimes (3 M_1 M_f - 3 M_f \overline{M}_1)
\\ &
- \iu (\Gamma_{453} + \Gamma_{783} + \Gamma_{693}) \otimes 
(M_1 M_p - \iu M_p \overline{M}_1 + 2 M_1 M_c - 2 M_c \overline{M}_1)
\\ & 
+ (\Gamma_{450} + \Gamma_{780} + \Gamma_{690}) \otimes 
(M_1 M_s - M_s \overline{M}_1 + 2 M_1 M_f - 2 M_f M_1)
\\ & 
- \iu (\Gamma_{01245} + \Gamma_{01278} + \Gamma_{01269}) 
\otimes (M_1 M_a' - M_a' \overline{M}_1 
+ 2 M_1 M_b - 2 M_b \overline{M}_1)
\\ & 
- (\Gamma_{31245} + \Gamma_{31278} + \Gamma_{31269}) 
\otimes (M_1 M_a - M_a \overline{M}_1 
+ 2 M_1 M_b' - 2 M_b' \overline{M}_1)
\\ & 
- \tfrac{1}{8} \iu (\Gamma_1 {-} \iu \Gamma_2) \Gamma_3 
(\Gamma_4 {-} \iu \Gamma_5)(\Gamma_6 {-} \iu \Gamma_9)
(\Gamma_7 {-} \iu \Gamma_8) \otimes (- 3 M_1 M_2 - 3 M_2 \overline{M}_1)~, 
\end{align*}
where ($h.c$ denotes the Hermitian conjugate of the preceding term)
\begin{align*}
\mathcal{Z}_1 &= (M_s M_p^\dagger + M_a' M_a^\dagger + 3 M_b' M_b^\dagger 
+ 3 M_c M_f^\dagger + \tfrac{1}{8} M_2 M_2^\dagger) + \mathrm{h.c}\,,
\\ 
\mathcal{Z}_2 &= (M_s M_a'{}^\dagger + M_p M_a^\dagger + 3 M_b' M_c^\dagger 
+ 3 M_b M_f^\dagger - \tfrac{1}{8} M_2 M_2^\dagger) + \mathrm{h.c}\,,
\\ 
\mathcal{Z}_3 &= (M_1 M_1^\dagger + \tfrac{1}{8} M_2 M_2^\dagger  
+ M_s M_f^\dagger + M_p M_c^\dagger + M_a' M_b^\dagger \\ & {}\quad 
+ M_a M_b'{}^\dagger + M_b' M_b'{}^\dagger + M_b M_b^\dagger 
+ M_c M_c^\dagger + M_f M_f^\dagger ) + \mathrm{h.c}\,,
\\ 
\mathcal{Z}_4 &= (M_s M_b^\dagger {+} M_p M_b'{}^\dagger 
{+} M_a' M_f^\dagger {+} M_a M_c^\dagger {+} 2 M_b' M_c^\dagger {+} 2
M_b M_f^\dagger  {-} \tfrac{1}{8} M_2 M_2^\dagger) + \mathrm{h.c}\,,
\\ 
\mathcal{Z}_5 &= (M_s M_c{}^\dagger {+} M_p M_f^\dagger 
{+} M_a' M_b'{}^\dagger {+} M_a M_b^\dagger {+} 2 M_b' M_b^\dagger 
{+} 2 M_c M_f^\dagger {+} \tfrac{1}{8} M_2 M_2^\dagger) + \mathrm{h.c}\,,
\\
\mathcal{Z}_6 &= (M_s M_a^\dagger + M_p M_a'{}^\dagger + 3 M_b' M_f^\dagger 
+ 3 M_b M_c^\dagger - \tfrac{1}{8} M_2 M_2^\dagger) + \mathrm{h.c}\,,
\\ 
\mathcal{Z}_7 &= \iu (((M_s + M_p + M_a' + M_a + 3 M_b' + 3 M_b 
+ 3 M_c + 3 M_f) M_2^\dagger) - \mathrm{h.c})\,,
\\
\mathcal{Z}_8 &= ((M_s + M_p + M_a' + M_a + 3 M_b' + 3 M_b 
+ 3 M_c + 3 M_f) M_2^\dagger) + \mathrm{h.c}\,.
\end{align*}

This gives 
\begin{align*}
\hat{\sigma}(\eta) &= \mathcal{P} \Big( \epsilon \one_4 \otimes 
{\arraycolsep 0pt
\left( \begin{array}{cccc} 
\hat{\sigma}(\eta_{(1)}) & 0 & 0 & 
\hat{\sigma}(\eta_{(2)})  \\
0 & \hat{\sigma}(\eta_{(1)}) ~{}& 
\hat{\sigma}(\eta_{(2)}) & 0 \\
0 & \hat{\sigma}(\eta_{(2)} )^\dagger ~{} & 
B \overline{\hat{\sigma}(\eta_{(1)})} B & 0 \\
\hat{\sigma}(\eta_{(2)})^\dagger & 0 & 0 & 
B \overline{\hat{\sigma}(\eta_{(1)})} B 
\end{array} \right) \Big) \mathcal{P}\,,
}
\\
\hat{\sigma}(\eta_{(1)}) &= 
- \iu \Theta \otimes (M_s M_b'{}^\dagger + M_b' M_s^\dagger 
+ M_p M_b^\dagger + M_b M_p^\dagger 
+ M_a' M_c^\dagger + M_c M_a'{}^\dagger \\ & {}\quad 
{+} M_a M_f^\dagger {+} M_f M_a^\dagger {+} 2 M_b' M_f^\dagger 
{+} 2 M_f M_b'{}^\dagger {+} 2 M_b M_c^\dagger {+} 2 M_c M_b^\dagger 
{-} \tfrac{1}{4} M_2 M_2^\dagger)
\\
\hat{\sigma}(\eta_{(2)}) &= 
\iu \Upsilon_1 \otimes (3 M_1 M_b' - 3 M_b' \overline{M}_1) 
+ \Upsilon_2 \otimes (3 M_1 M_b - 3 M_b \overline{M}_1) 
\\ & 
+ \Phi_1 \otimes (3 M_1 M_c - 3 M_c \overline{M}_1) 
- \iu \Phi_2 \otimes (3 M_1 M_f - 3 M_f \overline{M}_1)
\\ &
+ \Phi_3 \otimes (M_1 M_s - M_s \overline{M}_1 
+ 2 M_1 M_f - 2 M_f \overline{M}_1)
\\ & 
- \iu \Phi_4 \otimes (M_1 M_p - M_p \overline{M}_1 
+ 2 M_1 M_c - 2 M_c \overline{M}_1)
\\ & 
- \iu \Psi_1 \otimes (M_1 M_a' - M_a' \overline{M}_1 
+ 2 M_1 M_b - 2 M_b \overline{M}_1)
\\ & 
- \Psi_2 \otimes (M_1 M_a - M_a \overline{M}_1 
+ 2 M_1 M_b' - 2 M_b' \overline{M}_1)
\\ & 
- \iu \Psi_3 \otimes (- 3 M_1 M_2 - 3 M_2 \overline{M}_1)
\end{align*}
and
\begin{align*}
\mathcal{J}^2 &= \textstyle \sum([\pi(\mathfrak{g}),[\dots,
[\pi(\mathfrak{g}), Y^2_\perp]]\dots] 
+ \{\pi(\mathfrak{g}),\pi(\mathfrak{g})\}) \\
&=\left( \begin{array}{cccc} 
P_+ \mathcal{J}_{(1)} P_+ & 0 & 0 & 0 \\
0 & P_+ \mathcal{J}_{(1)} P_+ & 0 & 0 \\
0 & 0 & P_- B \overline{\mathcal{J}_{(1)}} B P_- & 0 \\
0 & 0 & 0 & P_- B \overline{\mathcal{J}_{(1)}} B P_-
\end{array} \right) ,
\\
\mathcal{J}_{(1)} & = C^\infty(M) \otimes \boldsymbol{1} 
\otimes \mathbb{C} \one_3
+ C^\infty(M) \otimes \iu \boldsymbol{45} 
\otimes (\mathbb{C} \mathcal{Z}_1 + \mathbb{C} \mathcal{Z}_2)
\\ &
+ C^\infty(M) \otimes \boldsymbol{210} \otimes (\mathbb{C} \mathcal{Z}_3 
+ \mathbb{C} \mathcal{Z}_4 + \mathbb{C} \mathcal{Z}_5 
+ \mathbb{C} \mathcal{Z}_6 + \mathbb{C} \mathcal{Z}_7 
+ \mathbb{C} \mathcal{Z}_8 + \mathbb{C} \one_3)~.
\end{align*}

The field strength $\mathcal{F}$ of the gauge potential $\rho=A+
\pi(\eta)$ is given in \eqref{F}. Let $\theta$ be the spacetime 0-form
component of $\mathcal{F}_\perp$ orthogonal to $\mathcal{J}^2$, as
given in \eqref{th}. Introducing 
\begin{align*}
\tilde{\Theta} &= \Theta + (\Gamma_{45} + \Gamma_{78} +
\Gamma_{69})~, \qquad{} 
\tilde{\Upsilon}_1 = \Upsilon_1 + \Gamma_0 ~, \qquad{} 
\tilde{\Upsilon}_2 = \Upsilon_2 + \Gamma_3 ~, \hspace*{-\textwidth} 
\\
\tilde{\Phi}_1 &= \Phi_1 + \Gamma_{120} ~, &
\tilde{\Phi}_2 &= \Phi_2 + \Gamma_{123} ~, 
\\
\tilde{\Phi}_3 &= \Phi_3 + (\Gamma_{453} + \Gamma_{783} + \Gamma_{693}) ~, &
\tilde{\Phi}_4 &= \Phi_4 + (\Gamma_{450} + \Gamma_{780} + \Gamma_{690}) ~, 
\\
\tilde{\Psi}_1 &= \Psi_1 
+ (\Gamma_{01245} + \Gamma_{01278} + \Gamma_{01269}) ~,&
\tilde{\Psi}_2 &= \Psi_2 
+ (\Gamma_{31245} + \Gamma_{31278} + \Gamma_{31269}) ~, 
\\
\tilde{\Psi}_3 &= \Psi_3 + \tfrac{1}{8} (\Gamma_1 {-} \iu \Gamma_2) 
\Gamma_3 (\Gamma_4 {-} \iu \Gamma_5)(\Gamma_6 {-} \iu \Gamma_9)
(\Gamma_7 {-} \iu \Gamma_8)~, \hspace*{-\textwidth}
\end{align*}
we find after straightforward but apparently lengthy calculation
\begin{align*}
\theta &= \epsilon \one_4 \otimes \left( \begin{array}{cccc} 
P_+ \theta_{(1)} P_+ & 0 & 0 & P_+ \theta_{(2)} P_- \\
0 & P_+ \theta_{(1)} P_+ & P_+ \theta_{(2)} P_- & 0 \\
0 & P_- \theta_{(2)}^\dagger P_+ & P_- B \overline{\theta_{(1)}} B
P_- & 0 \\
P_- \theta_{(2)}^\dagger P_+ & 0 & 0 & P_- B \overline{\theta_{(1)}}
B P_- \end{array} \right) ,
\\
\theta_{(1)} &= \textstyle
\sum_i \theta_{1}^i \otimes (Q_i^1)^\perp + 
\sum_j \theta_{45}^j \otimes (Q_j^{45})_\perp 
+ \sum_k \theta_{210}^k \otimes (Q_k^{210})_\perp 
\\ 
&= \tfrac{1}{2} (3\, \one_{32} - (\iu \tilde{\Theta})^2_1) 
\otimes \tilde{M}_{\{11\}} 
\\ &
+ \tfrac{1}{2} (\one_{32} -(\tilde{\Upsilon}^2_1)_1) 
\otimes \tilde{M}_{\{ss\}} 
+ \tfrac{1}{2} (\one_{32} -(\tilde{\Upsilon}^2_2)_1) 
\otimes \tilde{M}_{\{pp\}} 
\\ & 
+ \tfrac{1}{2} (\one_{32} -(\iu \tilde{\Phi}_1)^2_1) 
\otimes \tilde{M}_{\{a'a'\}} 
+ \tfrac{1}{2} (\one_{32} -(\iu \tilde{\Phi}_2)^2_1) 
\otimes \tilde{M}_{\{aa\}} 
\\ & 
+ \tfrac{1}{2} (3\, \one_{32} -(\iu \tilde{\Phi}_3)^2_1) 
\otimes \tilde{M}_{\{b'b'\}} 
+ \tfrac{1}{2} (3\, \one_{32} -(\iu \tilde{\Phi}_4)^2_1) 
\otimes \tilde{M}_{\{bb\}} 
\\ & 
+ \tfrac{1}{2} (3\, \one_{32} -(\tilde{\Psi}_1\tilde{\Psi}_1^\dagger)_1) 
\otimes \tilde{M}_{\{cc\}} 
+ \tfrac{1}{2} (3\, \one_{32} -(\tilde{\Psi}_2\tilde{\Psi}_2^\dagger)_1) 
\otimes \tilde{M}_{\{ff\}}  
\\ & 
+ \tfrac{1}{2} (16\, \one_{32} -(\tilde{\Psi}_3 \tilde{\Psi}_3^\dagger)_1) 
\otimes \tilde{M}_{\{22\}}  
- (\tilde{\Upsilon}_1 \tilde{\Upsilon}_2)_1 \otimes \tilde{M}_{[sp]} 
\\ & 
+ (\tilde{\Phi}_1 \tilde{\Phi}_2)_1 \otimes \tilde{M}_{[a'a]} 
+ (\tilde{\Phi}_1 \tilde{\Phi}_3)_1 \otimes \tilde{M}_{\{a'b'\}} 
+ (\tilde{\Phi}_1 \tilde{\Phi}_4)_1 \otimes \tilde{M}_{[a'b]} 
\\ &
- (\tilde{\Phi}_2 \tilde{\Phi}_3)_1 \otimes \tilde{M}_{[ab']} 
+ (\tilde{\Phi}_2 \tilde{\Phi}_4)_1 \otimes \tilde{M}_{\{ab\}} 
+ (\tilde{\Phi}_3 \tilde{\Phi}_4)_1 \otimes \tilde{M}_{[b'b]} 
\\ &
- \tfrac{1}{2} \iu (\tilde{\Psi}_1 \tilde{\Psi}_2^\dagger 
- \tilde{\Psi}_2 \tilde{\Psi}_1^\dagger)_1 \otimes \tilde{M}_{\{cf\}} 
- \tfrac{1}{2} (\tilde{\Psi}_1 \tilde{\Psi}_2^\dagger 
+ \tilde{\Psi}_2 \tilde{\Psi}_1^\dagger)_1 \otimes \tilde{M}_{[cf]} 
\\ &
+ \tfrac{1}{2} \iu (\tilde{\Psi}_1 \tilde{\Psi}_3^\dagger 
- \tilde{\Psi}_3 \tilde{\Psi}_1^\dagger)_1 \otimes \tilde{M}_{[c2]} 
- \tfrac{1}{2} (\tilde{\Psi}_1 \tilde{\Psi}_3^\dagger 
+ \tilde{\Psi}_3 \tilde{\Psi}_1^\dagger)_1 \otimes \tilde{M}_{\{c2\}} 
\\ & 
+ \tfrac{1}{2} \iu (\tilde{\Psi}_2 \tilde{\Psi}_3^\dagger 
- \tilde{\Psi}_3 \tilde{\Psi}_2^\dagger)_1 \otimes \tilde{M}_{\{f2\}} 
+ \tfrac{1}{2} (\tilde{\Psi}_2 \tilde{\Psi}_3^\dagger 
+ \tilde{\Psi}_3 \tilde{\Psi}_2^\dagger)_1 \otimes \tilde{M}_{[f2]}  
\\[1ex] &
- \iu (\tilde{\Upsilon}_1 \tilde{\Upsilon}_2)_{45} \otimes M_{\{sp\}}  
\\ &
+ \iu (\tilde{\Upsilon}_1 \tilde{\Phi}_1)_{45} \otimes M_{\{sa'\}} 
+ \iu (\tilde{\Upsilon}_1 \tilde{\Phi}_2)_{45} \otimes M_{[sa]} 
+ \iu (\tilde{\Upsilon}_1 \tilde{\Phi}_3 -\tilde{\Theta})_{45} 
\otimes M_{\{sb'\}}  
\\ & 
+ \iu (\tilde{\Upsilon}_1 \tilde{\Phi}_4)_{45} \otimes M_{[sb]}  
- \iu (\tilde{\Upsilon}_2 \tilde{\Phi}_1)_{45} \otimes M_{[pa']}  
+ \iu (\tilde{\Upsilon}_2 \tilde{\Phi}_2)_{45} \otimes M_{\{pa\}}  
\\ & 
- \iu (\tilde{\Upsilon}_2 \tilde{\Phi}_3)_{45} \otimes M_{[pb']}  
+ \iu (\tilde{\Upsilon}_2 \tilde{\Phi}_4 -\tilde{\Theta})_{45} 
\otimes M_{\{pb\}}  
\\ & 
+ \iu (\tilde{\Phi}_1 \tilde{\Phi}_2)_{45} \otimes M_{\{a'a\}}  
- \iu (\tilde{\Phi}_1 \tilde{\Phi}_3)_{45} \otimes M_{[a'b']} 
+ \iu (\tilde{\Phi}_1 \tilde{\Phi}_4)_{45} \otimes M_{\{a'b\}}  
\\ & 
- \tfrac{1}{2} \iu (\tilde{\Phi}_1 \tilde{\Psi}_1^\dagger 
+ \tilde{\Psi}_1 \tilde{\Phi}_1 + 2 \tilde{\Theta})_{45} \otimes M_{\{a'c\}}  
- \tfrac{1}{2} (\tilde{\Phi}_1 \tilde{\Psi}_1^\dagger 
- \tilde{\Psi}_1 \tilde{\Phi}_1)_{45} \otimes M_{[a'c]} 
\\ & 
- \tfrac{1}{2} \iu (\tilde{\Phi}_1 \tilde{\Psi}_2^\dagger 
+ \tilde{\Psi}_2 \tilde{\Phi}_1)_{45} \otimes M_{[a'f]}
+ \tfrac{1}{2} (\tilde{\Phi}_1 \tilde{\Psi}_2^\dagger 
- \tilde{\Psi}_2 \tilde{\Phi}_1)_{45} \otimes M_{\{a'f\}} 
\\ & 
- \tfrac{1}{2} \iu (\tilde{\Phi}_1 \tilde{\Psi}_3^\dagger 
+ \tilde{\Psi}_3 \tilde{\Phi}_1)_{45} \otimes M_{\{a'2\}}
- \tfrac{1}{2} (\tilde{\Phi}_1 \tilde{\Psi}_3^\dagger 
- \tilde{\Psi}_3 \tilde{\Phi}_1)_{45} \otimes M_{[a'2]} 
\\ &
- \iu (\tilde{\Phi}_2 \tilde{\Phi}_3)_{45} \otimes M_{\{ab'\}}  
- \iu (\tilde{\Phi}_2 \tilde{\Phi}_4)_{45} \otimes M_{[ab]}
+ \iu (\tilde{\Phi}_3 \tilde{\Phi}_4)_{45} \otimes M_{\{b'b\}} 
\\ & 
+ \tfrac{1}{2} \iu (\tilde{\Phi}_2 \tilde{\Psi}_1^\dagger 
+ \tilde{\Psi}_1 \tilde{\Phi}_2)_{45} \otimes M_{[ac]}
- \tfrac{1}{2} (\tilde{\Phi}_2 \tilde{\Psi}_1^\dagger 
- \tilde{\Psi}_1 \tilde{\Phi}_2)_{45} \otimes M_{\{ac\}}
\\ & 
- \tfrac{1}{2} \iu (\tilde{\Phi}_2 \tilde{\Psi}_2^\dagger 
+ \tilde{\Psi}_2 \tilde{\Phi}_2 + 2 \tilde{\Theta})_{45} \otimes M_{\{af\}}
- \tfrac{1}{2} (\tilde{\Phi}_2 \tilde{\Psi}_2^\dagger 
- \tilde{\Psi}_2 \tilde{\Phi}_2)_{45} \otimes M_{[af]}
\\ & 
+ \tfrac{1}{2} \iu (\tilde{\Phi}_2 \tilde{\Psi}_3^\dagger 
+ \tilde{\Psi}_3 \tilde{\Phi}_2)_{45} \otimes M_{[a2]}
- \tfrac{1}{2} (\tilde{\Phi}_2 \tilde{\Psi}_3^\dagger 
- \tilde{\Psi}_3 \tilde{\Phi}_2)_{45} \otimes M_{\{a2\}}
\\ & 
- \tfrac{1}{2} \iu (\tilde{\Phi}_3 \tilde{\Psi}_1^\dagger 
+ \tilde{\Psi}_1 \tilde{\Phi}_3)_{45} \otimes M_{\{b'c\}} 
- \tfrac{1}{2} (\tilde{\Phi}_3 \tilde{\Psi}_1^\dagger 
- \tilde{\Psi}_1 \tilde{\Phi}_3)_{45} \otimes M_{[b'c]} 
\\ & 
- \tfrac{1}{2} \iu (\tilde{\Phi}_3 \tilde{\Psi}_2^\dagger 
+ \tilde{\Psi}_2 \tilde{\Phi}_3)_{45} \otimes M_{[b'f]}
+ \tfrac{1}{2} (\tilde{\Phi}_3 \tilde{\Psi}_2^\dagger 
- \tilde{\Psi}_2 \tilde{\Phi}_3 - 4 \iu \tilde{\Theta})_{45} 
\otimes M_{\{b'f\}} 
\\ & 
- \tfrac{1}{2} \iu (\tilde{\Phi}_3 \tilde{\Psi}_3^\dagger 
+ \tilde{\Psi}_3 \tilde{\Phi}_3)_{45} \otimes M_{\{b'2\}}
- \tfrac{1}{2} (\tilde{\Phi}_3 \tilde{\Psi}_3^\dagger 
- \tilde{\Psi}_3 \tilde{\Phi}_3)_{45} \otimes M_{[b'2]}
\\ & 
+ \tfrac{1}{2} \iu (\tilde{\Phi}_4 \tilde{\Psi}_1^\dagger 
+ \tilde{\Psi}_1 \tilde{\Phi}_4)_{45} \otimes M_{[bc]}
- \tfrac{1}{2} (\tilde{\Phi}_4 \tilde{\Psi}_1^\dagger 
- \tilde{\Psi}_1 \tilde{\Phi}_4 + 4 \iu \tilde{\Theta})_{45} 
\otimes M_{\{bc\}} 
\\ & 
- \tfrac{1}{2} \iu (\tilde{\Phi}_4 \tilde{\Psi}_2^\dagger 
+ \tilde{\Psi}_2 \tilde{\Phi}_4)_{45} \otimes M_{\{bf\}}
- \tfrac{1}{2} (\tilde{\Phi}_4 \tilde{\Psi}_2^\dagger 
- \tilde{\Psi}_2 \tilde{\Phi}_4)_{45} \otimes M_{[bf]}
\\ & 
+ \tfrac{1}{2} \iu (\tilde{\Phi}_4 \tilde{\Psi}_3^\dagger 
+ \tilde{\Psi}_3 \tilde{\Phi}_4)_{45} \otimes M_{[b2]}
- \tfrac{1}{2} (\tilde{\Phi}_4 \tilde{\Psi}_3^\dagger 
- \tilde{\Psi}_3 \tilde{\Phi}_4)_{45} \otimes M_{\{b2\}}
\\ & 
- \tfrac{1}{2} (\tilde{\Psi}_1 \tilde{\Psi}_1^\dagger)_{45} 
\otimes M_{\{cc\}} 
- \tfrac{1}{2} (\tilde{\Psi}_2 \tilde{\Psi}_2^\dagger)_{45} 
\otimes M_{\{ff\}} 
- \tfrac{1}{2} ((\tilde{\Psi}_3 \tilde{\Psi}_3^\dagger)_{45} 
- 16 \iu \tilde{\Theta}) \otimes M_{\{22\}} 
\\ &
- \tfrac{1}{2} \iu (\tilde{\Psi}_1 \tilde{\Psi}_2^\dagger 
- \tilde{\Psi}_2 \tilde{\Psi}_1^\dagger)_{45} \otimes M_{\{cf\}}
- \tfrac{1}{2} (\tilde{\Psi}_1 \tilde{\Psi}_2^\dagger 
+ \tilde{\Psi}_2 \tilde{\Psi}_1^\dagger)_{45} \otimes M_{[cf]} 
\\ & 
+ \tfrac{1}{2} \iu (\tilde{\Psi}_1 \tilde{\Psi}_3^\dagger 
- \tilde{\Psi}_3 \tilde{\Psi}_1^\dagger)_{45} \otimes M_{[c2]}
- \tfrac{1}{2} (\tilde{\Psi}_1 \tilde{\Psi}_3^\dagger 
+ \tilde{\Psi}_3 \tilde{\Psi}_1^\dagger)_{45} \otimes M_{\{c2\}} 
\\ & 
+ \tfrac{1}{2} \iu (\tilde{\Psi}_2 \tilde{\Psi}_3^\dagger 
- \tilde{\Psi}_3 \tilde{\Psi}_2^\dagger)_{45} \otimes M_{\{f2\}}
+ \tfrac{1}{2} (\tilde{\Psi}_2 \tilde{\Psi}_3^\dagger 
+ \tilde{\Psi}_3 \tilde{\Psi}_2^\dagger)_{45} \otimes M_{[f2]} 
\\[1ex]
& - \tfrac{1}{2} (\iu \tilde{\Theta})^2_{210} \otimes \hat{M}_{\{11\}}
\\ & 
+ (\tilde{\Upsilon}_1 \tilde{\Phi}_1)_{210} \otimes \hat{M}_{[sa']}
- (\tilde{\Upsilon}_1 \tilde{\Phi}_2)_{210} \otimes \hat{M}_{\{sa\}}
+ (\tilde{\Upsilon}_1 \tilde{\Phi}_3)_{210} \otimes \hat{M}_{[sb']}
\\ & 
- (\tilde{\Upsilon}_1 \tilde{\Phi}_4)_{210} \otimes \hat{M}_{\{sb\}}
+ (\tilde{\Upsilon}_2 \tilde{\Phi}_1)_{210} \otimes \hat{M}_{\{pa'\}}
+ (\tilde{\Upsilon}_2 \tilde{\Phi}_2)_{210} \otimes \hat{M}_{[pa]}
\\ &
+ (\tilde{\Upsilon}_2 \tilde{\Phi}_3)_{210} \otimes \hat{M}_{\{pb'\}}
+ (\tilde{\Upsilon}_2 \tilde{\Phi}_4)_{210} \otimes \hat{M}_{[pb]}
\\ & 
+ \tfrac{1}{2} (\tilde{\Upsilon}_1 \tilde{\Psi}_1^\dagger 
+ \tilde{\Psi}_1 \tilde{\Upsilon}_1)_{210} \otimes \hat{M}_{\{sc\}}
- \tfrac{1}{2} \iu (\tilde{\Upsilon}_1 \tilde{\Psi}_1^\dagger 
- \tilde{\Psi}_1 \tilde{\Upsilon}_1)_{210} \otimes \hat{M}_{[sc]}
\\ &
+ \tfrac{1}{2} (\tilde{\Upsilon}_1 \tilde{\Psi}_2^\dagger 
+ \tilde{\Psi}_2 \tilde{\Upsilon}_1)_{210} \otimes \hat{M}_{[sf]}
- \tfrac{1}{2} \iu (\tilde{\Upsilon}_1 \tilde{\Psi}_2^\dagger 
- \tilde{\Psi}_2 \tilde{\Upsilon}_1)_{210} \otimes \hat{M}_{\{sf\}}
\\ &
+ \tfrac{1}{2} (\tilde{\Upsilon}_1 \tilde{\Psi}_3^\dagger 
+ \tilde{\Psi}_3 \tilde{\Upsilon}_1)_{210} \otimes \hat{M}_{\{s2\}}
- \tfrac{1}{2} \iu (\tilde{\Upsilon}_1 \tilde{\Psi}_3^\dagger 
- \tilde{\Psi}_3 \tilde{\Upsilon}_1)_{210} \otimes \hat{M}_{[s2]}
\\ & 
- \tfrac{1}{2} (\tilde{\Upsilon}_2 \tilde{\Psi}_1^\dagger 
+ \tilde{\Psi}_1 \tilde{\Upsilon}_2)_{210} \otimes \hat{M}_{[pc]}
- \tfrac{1}{2} \iu (\tilde{\Upsilon}_2 \tilde{\Psi}_1^\dagger 
- \tilde{\Psi}_1 \tilde{\Upsilon}_2)_{210} \otimes \hat{M}_{\{pc\}}
\\ &
+ \tfrac{1}{2} (\tilde{\Upsilon}_2 \tilde{\Psi}_2^\dagger 
+ \tilde{\Psi}_2 \tilde{\Upsilon}_2)_{210} \otimes \hat{M}_{\{pf\}}
- \tfrac{1}{2} \iu (\tilde{\Upsilon}_2 \tilde{\Psi}_2^\dagger 
- \tilde{\Psi}_2 \tilde{\Upsilon}_2)_{210} \otimes \hat{M}_{[pf]}
\\ &
- \tfrac{1}{2} (\tilde{\Upsilon}_2 \tilde{\Psi}_3^\dagger 
+ \tilde{\Psi}_3 \tilde{\Upsilon}_2)_{210} \otimes \hat{M}_{[p2]}
- \tfrac{1}{2} \iu (\tilde{\Upsilon}_2 \tilde{\Psi}_3^\dagger 
- \tilde{\Psi}_3 \tilde{\Upsilon}_2)_{210} \otimes \hat{M}_{\{p2\}}
\\ &
+ \tfrac{1}{2} (\tilde{\Phi}_1^2)_{210} \otimes \hat{M}_{\{a'a'\}}
+ \tfrac{1}{2} (\tilde{\Phi}_2^2)_{210} \otimes \hat{M}_{\{aa\}}
\\ & 
+ \tfrac{1}{2} (\tilde{\Phi}_3^2)_{210} \otimes \hat{M}_{\{b'b'\}}
+ \tfrac{1}{2} (\tilde{\Phi}_4^2)_{210} \otimes \hat{M}_{\{bb\}}
\\ &
+ (\tilde{\Phi}_1 \tilde{\Phi}_2)_{210} \otimes \hat{M}_{[a'a]}
+ (\iu \tilde{\Phi}_1 \tilde{\Phi}_2 \Gamma_{11})_{210} 
  \otimes \hat{M}_{\{a'a\}}
+ (\tilde{\Phi}_1 \tilde{\Phi}_3)_{210} \otimes \hat{M}_{\{a'b'\}}
\\ &
- (\iu \tilde{\Phi}_1 \tilde{\Phi}_3 \Gamma_{11})_{210} 
\otimes \hat{M}_{[a'b']}
+ (\tilde{\Phi}_1 \tilde{\Phi}_4)_{210} \otimes \hat{M}_{[a'b]}
+ (\iu \tilde{\Phi}_1 \tilde{\Phi}_4 \Gamma_{11})_{210} 
  \otimes \hat{M}_{\{a'b\}}
\\ &
- (\tilde{\Phi}_2 \tilde{\Phi}_3)_{210} \otimes \hat{M}_{[ab']}
- (\iu \tilde{\Phi}_2 \tilde{\Phi}_3 \Gamma_{11})_{210} 
  \otimes \hat{M}_{\{ab'\}}
+ (\tilde{\Phi}_2 \tilde{\Phi}_4)_{210} \otimes \hat{M}_{\{ab\}}
\\ &
- (\iu \tilde{\Phi}_2 \tilde{\Phi}_4 \Gamma_{11})_{210} 
\otimes \hat{M}_{[ab]}
+ (\tilde{\Phi}_3 \tilde{\Phi}_4)_{210} \otimes \hat{M}_{[b'b]}
+ (\iu \tilde{\Phi}_3 \tilde{\Phi}_4 \Gamma_{11})_{210} 
  \otimes \hat{M}_{\{b'b\}}
\\ &
- \tfrac{1}{2} (\tilde{\Phi}_1 \tilde{\Psi}_1^\dagger 
- \tilde{\Psi}_1 \tilde{\Phi}_1)_{210} \otimes \hat{M}_{[a'c]}
- \tfrac{1}{2} \iu (\tilde{\Phi}_1 \tilde{\Psi}_1^\dagger
+ \tilde{\Psi}_1 \tilde{\Phi}_1)_{210} \otimes \hat{M}_{\{a'c\}}
\\ &
+ \tfrac{1}{2} (\tilde{\Phi}_1 \tilde{\Psi}_2^\dagger 
- \tilde{\Psi}_2 \tilde{\Phi}_1)_{210} \otimes \hat{M}_{\{a'f\}}
- \tfrac{1}{2} \iu (\tilde{\Phi}_1 \tilde{\Psi}_2^\dagger
+ \tilde{\Psi}_2 \tilde{\Phi}_1)_{210} \otimes \hat{M}_{[a'f]}
\\ &
- \tfrac{1}{2} (\tilde{\Phi}_1 \tilde{\Psi}_3^\dagger 
- \tilde{\Psi}_3 \tilde{\Phi}_1)_{210} \otimes \hat{M}_{[a'2]}
- \tfrac{1}{2} \iu (\tilde{\Phi}_1 \tilde{\Psi}_3^\dagger
+ \tilde{\Psi}_3 \tilde{\Phi}_1)_{210} \otimes \hat{M}_{\{a'2\}}
\\ &
- \tfrac{1}{2} (\tilde{\Phi}_2 \tilde{\Psi}_1^\dagger 
- \tilde{\Psi}_1 \tilde{\Phi}_2)_{210} \otimes \hat{M}_{\{ac\}}
+ \tfrac{1}{2} \iu (\tilde{\Phi}_2 \tilde{\Psi}_1^\dagger
+ \tilde{\Psi}_1 \tilde{\Phi}_2)_{210} \otimes \hat{M}_{[ac]}
\\ &
- \tfrac{1}{2} (\tilde{\Phi}_2 \tilde{\Psi}_2^\dagger 
- \tilde{\Psi}_2 \tilde{\Phi}_2)_{210} \otimes \hat{M}_{[af]}
- \tfrac{1}{2} \iu (\tilde{\Phi}_2 \tilde{\Psi}_2^\dagger
+ \tilde{\Psi}_2 \tilde{\Phi}_2)_{210} \otimes \hat{M}_{\{af\}}
\\ &
- \tfrac{1}{2} (\tilde{\Phi}_2 \tilde{\Psi}_3^\dagger 
- \tilde{\Psi}_3 \tilde{\Phi}_2)_{210} \otimes \hat{M}_{\{a2\}}
+ \tfrac{1}{2} \iu (\tilde{\Phi}_2 \tilde{\Psi}_3^\dagger
+ \tilde{\Psi}_3 \tilde{\Phi}_2)_{210} \otimes \hat{M}_{[a2]}
\\ &
- \tfrac{1}{2} (\tilde{\Phi}_3 \tilde{\Psi}_1^\dagger 
- \tilde{\Psi}_1 \tilde{\Phi}_3)_{210} \otimes \hat{M}_{[b'c]}
- \tfrac{1}{2} \iu (\tilde{\Phi}_3 \tilde{\Psi}_1^\dagger
+ \tilde{\Psi}_1 \tilde{\Phi}_3)_{210} \otimes \hat{M}_{\{b'c\}}
\\ &
+ \tfrac{1}{2} (\tilde{\Phi}_3 \tilde{\Psi}_2^\dagger 
- \tilde{\Psi}_2 \tilde{\Phi}_3)_{210} \otimes \hat{M}_{\{b'f\}}
- \tfrac{1}{2} \iu (\tilde{\Phi}_3 \tilde{\Psi}_2^\dagger
+ \tilde{\Psi}_2 \tilde{\Phi}_3)_{210} \otimes \hat{M}_{[b'f]}
\\ &
- \tfrac{1}{2} (\tilde{\Phi}_3 \tilde{\Psi}_3^\dagger 
- \tilde{\Psi}_3 \tilde{\Phi}_3)_{210} \otimes \hat{M}_{[b'2]}
- \tfrac{1}{2} \iu (\tilde{\Phi}_3 \tilde{\Psi}_3^\dagger
+ \tilde{\Psi}_3 \tilde{\Phi}_3)_{210} \otimes \hat{M}_{\{b'2\}}
\\ &
- \tfrac{1}{2} (\tilde{\Phi}_4 \tilde{\Psi}_1^\dagger 
- \tilde{\Psi}_1 \tilde{\Phi}_4)_{210} \otimes \hat{M}_{\{bc\}}
+ \tfrac{1}{2} \iu (\tilde{\Phi}_4 \tilde{\Psi}_1^\dagger
+ \tilde{\Psi}_1 \tilde{\Phi}_4)_{210} \otimes \hat{M}_{[bc]}
\\ &
- \tfrac{1}{2} (\tilde{\Phi}_4 \tilde{\Psi}_2^\dagger 
- \tilde{\Psi}_2 \tilde{\Phi}_4)_{210} \otimes \hat{M}_{[bf]}
- \tfrac{1}{2} \iu (\tilde{\Phi}_4 \tilde{\Psi}_2^\dagger
+ \tilde{\Psi}_2 \tilde{\Phi}_4)_{210} \otimes \hat{M}_{\{bf\}}
\\ &
- \tfrac{1}{2} (\tilde{\Phi}_4 \tilde{\Psi}_3^\dagger 
- \tilde{\Psi}_3 \tilde{\Phi}_4)_{210} \otimes \hat{M}_{\{b2\}}
+ \tfrac{1}{2} \iu (\tilde{\Phi}_4 \tilde{\Psi}_3^\dagger
+ \tilde{\Psi}_3 \tilde{\Phi}_4)_{210} \otimes \hat{M}_{[b2]}
\\ &
- (\tilde{\Psi}_1 \tilde{\Psi}_1^\dagger)_{210} \otimes \hat{M}_{\{cc\}}
- (\tilde{\Psi}_2 \tilde{\Psi}_2^\dagger)_{210} \otimes \hat{M}_{\{ff\}}
- (\tilde{\Psi}_3 \tilde{\Psi}_3^\dagger)_{210} \otimes \hat{M}_{\{22\}}
\\ &
- \tfrac{1}{2} (\tilde{\Psi}_1 \tilde{\Psi}_2^\dagger 
+ \tilde{\Psi}_2 \tilde{\Psi}_1^\dagger)_{210} \otimes \hat{M}_{[cf]}
- \tfrac{1}{2} \iu (\tilde{\Psi}_1 \tilde{\Psi}_2^\dagger 
- \tilde{\Psi}_2 \tilde{\Psi}_1^\dagger)_{210} \otimes \hat{M}_{\{cf\}}
\\ & 
- \tfrac{1}{2} (\tilde{\Psi}_1 \tilde{\Psi}_3^\dagger 
+ \tilde{\Psi}_3 \tilde{\Psi}_1^\dagger)_{210} \otimes \hat{M}_{\{c2\}}
+ \tfrac{1}{2} \iu (\tilde{\Psi}_1 \tilde{\Psi}_3^\dagger 
- \tilde{\Psi}_3 \tilde{\Psi}_1^\dagger)_{210} \otimes \hat{M}_{[c2]}
\\ & 
+ \tfrac{1}{2} (\tilde{\Psi}_2 \tilde{\Psi}_3^\dagger 
+ \tilde{\Psi}_3 \tilde{\Psi}_2^\dagger)_{210} \otimes \hat{M}_{[f2]}
+ \tfrac{1}{2} \iu (\tilde{\Psi}_2 \tilde{\Psi}_3^\dagger 
- \tilde{\Psi}_3 \tilde{\Psi}_2^\dagger)_{210} \otimes \hat{M}_{\{f2\}}~,
\\[1.5ex]
\theta_{(2)} &= \textstyle 
\sum_i \theta_{10}^i \otimes Q_i^{10} + 
\sum_j \theta_{120}^j \otimes Q_j^{120} + 
\sum_k \theta_{126}^k \otimes Q_k^{126}
\\ 
&= - (\tilde{\Theta} \tilde{\Upsilon}_1)_{10} \otimes M_{\{1s\}}  
+ \iu (\tilde{\Theta} \tilde{\Upsilon}_2)_{10} \otimes M_{\{1p\}} 
+ (\tilde{\Theta} \tilde{\Phi}_1)_{10} \otimes M_{[1a']}
\\ & 
- \iu (\tilde{\Theta} \tilde{\Phi}_2)_{10} \otimes M_{[1a]}
+ ((\tilde{\Theta} \tilde{\Phi}_3)_{10} + 3 \tilde{\Upsilon}_1)
\otimes M_{[1b']} 
- \iu ((\tilde{\Theta} \tilde{\Phi}_4)_{10} + 3 \tilde{\Upsilon}_2)
\otimes M_{[1b]} 
\\[0.5ex] & 
+ ((\iu \tilde{\Theta} \tilde{\Upsilon}_1)_{120} {-} \iu \tilde{\Phi}_3)
\otimes M_{[1s]}
- \iu ((\iu \tilde{\Theta} \tilde{\Upsilon}_2)_{120} 
{-} \iu \tilde{\Phi}_4) \otimes M_{[1p]}
+ (\iu \tilde{\Theta} \tilde{\Phi}_1)_{120} \otimes M_{\{1a'\}}
\\ &
- \iu (\iu \tilde{\Theta} \tilde{\Phi}_2)_{120} \otimes M_{\{1a\}}
+ (\iu \tilde{\Theta} \tilde{\Phi}_3)_{120} \otimes M_{\{1b'\}} 
- \iu (\iu \tilde{\Theta} \tilde{\Phi}_4)_{120} \otimes M_{\{1b\}} 
\\ &
- ((\iu \tilde{\Theta} \tilde{\Psi}_1)_{120} + 3 \iu \tilde{\Phi}_1 
+ 2 \tilde{\Phi}_4) \otimes M_{[1c]}
+ \iu ((\iu \tilde{\Theta} \tilde{\Psi}_2)_{120} + 3 \iu \tilde{\Phi}_2
- 2 \tilde{\Phi}_3) \otimes M_{[1f]} \\ &
- (\iu \tilde{\Theta} \tilde{\Psi}_3)_{120} \otimes M_{[12]}
\\[0.5ex] &
+ ((\tilde{\Theta} \tilde{\Phi}_1)_{126} - \tilde{\Psi}_1) 
\otimes M_{[1a']}
- \iu ((\tilde{\Theta} \tilde{\Phi}_2)_{126} -\tilde{\Psi}_2)
\otimes M_{[1a]}
\\ & 
+ ((\tilde{\Theta} \tilde{\Phi}_3)_{126} + 2 \iu \tilde{\Psi}_2)
\otimes M_{[1b']} 
- \iu ((\tilde{\Theta} \tilde{\Phi}_4)_{126} - 2 \iu \tilde{\Psi}_1)
\otimes M_{[1b]} 
\\ & 
+ (\tilde{\Theta} \tilde{\Psi}_1)_{126} \otimes M_{\{1c\}}
- \iu (\tilde{\Theta} \tilde{\Psi}_2)_{126} \otimes M_{\{1f\}} 
+ ((\tilde{\Theta} \tilde{\Psi}_3)_{126} + 3 \iu \tilde{\Psi}_3)
\otimes M_{\{12\}}~,
\end{align*}
with
\begin{align*}
\tilde{M}_{\{\alpha\beta\}} &= (M_\alpha M_\beta^\dagger 
+ M_\beta M_\alpha^\dagger)^{(\perp)} ~, & 
\tilde{M}_{[\alpha\beta]} &= (\iu M_\alpha M_\beta^\dagger 
- \iu M_\beta M_\alpha^\dagger)^{(\perp)} ~, 
\\
M_{\{\alpha\beta\}} &= (M_\alpha M_\beta^\dagger 
+ M_\beta M_\alpha^\dagger)_\perp ~, & 
M_{[\alpha\beta]} &= (\iu M_\alpha M_\beta^\dagger 
- \iu M_\beta M_\alpha^\dagger)_\perp ~, & 
(\alpha{=}1 {\Rightarrow} \beta{=}1)
\\
\hat{M}_{\{\alpha\beta\}} &= (M_\alpha M_\beta^\dagger 
+ M_\beta M_\alpha^\dagger)^\perp ~, & 
\hat{M}_{[\alpha\beta]} &= (\iu M_\alpha M_\beta^\dagger 
- \iu M_\beta M_\alpha^\dagger)^\perp ~, 
\\
M_{\{1\alpha\}} &= (M_1 M_\alpha + M_\alpha \overline{M}_1) ~, & 
M_{[1\alpha]} &= (\iu M_1 M_\alpha - \iu M_\alpha \overline{M}_1) ~, 
& (\alpha \neq 1)
\\
(Q_i^1)^{(\perp)} &= Q_i - \tfrac{1}{3}\mathrm{tr}(Q_i) \one_3 ~,
\\
(Q_j^{45})_\perp &= Q_j^{45} 
- \textstyle \sum_{a,b=1}^2 \mathrm{tr}(Q_j^{45} \mathcal{Z}_a) 
\mathcal{T}_{ab} \mathcal{Z}_b~, \qquad
\sum_{a=1}^2 \mathcal{T}_{a'a} \mathrm{tr}(\mathcal{Z}_{a}\mathcal{Z}_{b})
= \delta_{a'b}~, \hspace*{-\textwidth}
\\
(Q_j^{210})^\perp &= Q_j^{210} 
- \textstyle \sum_{a,b=3}^9 \mathrm{tr}(Q_j^{210} \mathcal{Z}_a) 
\mathcal{T}_{ab} \mathcal{Z}_b~, \qquad
\sum_{a=3}^9 \mathcal{T}_{a'a} \mathrm{tr}(\mathcal{Z}_{a}\mathcal{Z}_{b})
= \delta_{a'b}~, \hspace*{-\textwidth}
\end{align*}
with $\mathcal{Z}_9=\one_3$. We have the following 8 constraints due to 
the $\mathcal{Z}_i$ of our ideal:
\begin{align}
0 &= M_{\{sp\}} + M_{\{a'a\}} + 3 M_{\{b'b\}} 
+ 3 M_{\{cf\}} + 8 M_{\{22\}}~, \notag
\\ 
0 &= M_{\{sa'\}} + M_{\{pa\}} + 3 M_{\{b'c\}} 
+ 3 M_{\{bf\}} - 8 M_{\{22\}}~, \notag
\\ 
0 &= \hat{M}_{\{11\}} + 8 \hat{M}_{\{22\}} 
+ \hat{M}_{\{sf\}} + \hat{M}_{\{pc\}} + \hat{M}_{\{a'b\}} 
+ \hat{M}_{\{ab'\}} \notag \\ & 
+ \hat{M}_{\{b'b'\}} + \hat{M}_{\{bb\}} 
+ \hat{M}_{\{cc\}} + \hat{M}_{\{ff\}} ~, \notag 
\\ 
0 &= \hat{M}_{\{sb\}} + \hat{M}_{\{pb'\}} + \hat{M}_{\{a'f\}} 
+ \hat{M}_{\{ac\}} + 2 \hat{M}_{\{b'c\}} + 2 \hat{M}_{\{bf\}} 
- 8 \hat{M}_{\{22\}}~, \label{cns}
\\ 
0 &= \hat{M}_{\{sc\}} + \hat{M}_{\{pf\}} + \hat{M}_{\{a'b'\}}  
+ \hat{M}_{\{ab\}} + 2 \hat{M}_{\{b'b\}} + 2 \hat{M}_{\{cf\}} 
+ 8 \hat{M}_{\{22\}}~, \notag
\\
0 &= \hat{M}_{\{sa\}} + \hat{M}_{\{pa'\}} + 3 \hat{M}_{\{b'f\}} 
+ 3 \hat{M}_{\{bc\}} - 8 \hat{M}_{\{22\}}~, \notag
\\ 
0 &= \hat{M}_{[s2]} + \hat{M}_{[p2]} + \hat{M}_{[a'2]} 
+ \hat{M}_{[a2]} + 3 \hat{M}_{[b'2]} + 3 \hat{M}_{[b2]} 
+ 3 \hat{M}_{[c2]} + 3 \hat{M}_{[f2]}~, \notag
\\
0 &= \hat{M}_{\{s2\}} + \hat{M}_{\{p2\}} + \hat{M}_{\{a'2\}} 
+ \hat{M}_{\{a2\}} + 3 \hat{M}_{\{b'2\}} + 3 \hat{M}_{\{b2\}} 
+ 3 \hat{M}_{\{c2\}} + 3 \hat{M}_{\{f2\}}~. \notag
\end{align}

According to the general theory we have to investigate whether or not 
the connection form $\rho$ receives an extra contribution
$\rho' \in \Lambda^1 \otimes \mathbf{r}^0 + \Lambda^0 \gamma^5 \otimes 
\mathbf{r}^1$. Due to the symmetries of our setting and the
requirement that the $\mathbf{r}^i$ commute with
$\hat{\pi}(\mathrm{so(10}))$, the matrices $\mathbf{r}^i$ have the
general form  
{\arraycolsep -1pt
\begin{align*}
\mathbf{r}^0 &= \left( \begin{array}{cccc}
\,\iu P_+ (\one_{32} \otimes M_0) P_+ \!\! & 0 & 0 & 0 \\
0 & \!\! \iu P_+ (\one_{32} \otimes M_0) P_+ \!\! & 0 & 0 \\
0 & 0 & \!\! - \iu P_- (\one_{32} \otimes \overline{M_0}) P_- \!\! & 0 \\
0 & 0 & 0 & \!\! - \iu P_- (\one_{32} \otimes \overline{M_0}) P_- \,
\end{array} \right), 
\\
\mathbf{r}^1 &= \left( \begin{array}{cccc} 
0 & P_+ (\one_{32} \otimes M_0') P_+ & 0 & 0 \\
\epsilon P_+ (\one_{32} \otimes M_0') P_+ & 0 & 0 & 0 \\
0 & 0 & 0 & \epsilon P_- (\one_{32} \otimes \overline{M_0'}) P_- \\
0 & 0 & P_- (\one_{32} \otimes \overline{M_0'}) P_- & 0 
\end{array} \right),
\end{align*}}%
with $M_0=M_0^\dagger$ and $M_0'=M_0'{}^\dagger$. The condition 
$[\mathbf{r}^0, \hat{\pi}(\eta)] \in
\hat{\pi}(\Omega^1(\mathfrak{a}))$ yields from the
$\boldsymbol{45}$-sector $\iu [M_0,M_1] \in \mathbb{R} M_1$, which
fixes $M_0$ up to three parameters. The $\boldsymbol{10}$-sector
yields $\iu (M_0 M_s + M_s \overline{M_0}) \in \mathbb{R} M_s 
+ \iu \mathbb{R} M_p$ and $\iu (M_0 M_p + M_p \overline{M_0}) 
\in \mathbb{R} M_p + \iu \mathbb{R} M_s$. The r.h.s.\ are
two-dimensional so that both $\iu (M_0 M_s + M_s \overline{M_0})$ and
$\iu (M_0 M_p + M_p \overline{M_0})$ are orthogonal to 7-dimensional
spaces. It is clear that there exists no solution for $M_0$ in
general. 

The condition $\{\mathbf{r}^1,\hat{\pi}(\eta)\} \in 
\hat{\pi}(\Omega^2(\mathfrak{a})) 
+ \{ \hat{\pi}(\mathfrak{a}),\hat{\pi}(\mathfrak{a})\}$ derived from 
$\{\rho',\pi(\omega^1)\} \in \pi(\Omega^2)$ gives from the 
$\boldsymbol{45}$-sector no condition at all, because 
$-\iu \Theta \otimes (M_0' M_1+M_1 M_0')$ is contained in $\theta_{(1)}$ for 
any $M_0'$. From the $\boldsymbol{10}$-sector we get $2$ times $(9-6)$ 
conditions, from the $\boldsymbol{120}$-sector $4$ times $(9-9)$ 
conditions and from the $\boldsymbol{10}$-sector $3$ times $(9-7)$
conditions. This means that there will not exist a solution for $M_0'$
in general. There are no extra contributions to the gauge potential
possible. 

In the same way one shows that the graded centralizer $\mathcal{C}^2$ is 
trivial, $\mathcal{C}^2 = C^\infty(M) \mathcal{P} \subset 
\{\pi(\mathfrak{g}),\pi(\mathfrak{g})\}$. There is also no extra 
contribution to the ideal $\mathcal{J}^2$.

\section{The bosonic action}
\label{bose}

The bosonic action \eqref{SB} is now given by
\[
S_B = \tfrac{\epsilon}{192 g^2} \int_M \mathrm{tr}(\mathcal{F}^2)\,dx 
= \int_M (\mathcal{L}_2 + \mathcal{L}_1 + V)\,dx ~,
\]
where $g$ is the $\mathrm{so(10)}$ coupling constant. In this formula, 
$V=\tfrac{\epsilon}{192 g^2} \mathrm{tr}(\theta^2)$ is the 
Higgs potential, which in more detail is given by
\begin{align*}
V &=\tfrac{\epsilon}{48 g^2} \big( 
\sum_{i,i'} \mathrm{tr}(P_+ \theta^i_1 \theta^{i'}_1)\, 
\mathrm{tr}((Q_i^1)^{(\perp)} (Q_{i'}^1)^{(\perp)} ) 
+ \sum_{j,j'} \mathrm{tr}(P_+ \theta^j_{45} \theta^{j'}_{45}) \,
\mathrm{tr}((Q_j^{45})_\perp (Q_{j'}^{45})_\perp ) 
\\
&+ \sum_{k,k'} \mathrm{tr}(P_+ \theta^k_{210} \theta^{k'}_{210}) \,
\mathrm{tr}((Q_k^{210})^\perp (Q_{k'}^{210})^\perp ) 
+ \sum_{k,k'} \mathrm{tr}(P_+ \theta^k_{126}
(\theta^{k'}_{126})^\dagger) \,
\mathrm{tr}(Q_k^{126} (Q_{k'}^{126})^\dagger) 
\\ 
& +  \sum_{i,i'} \mathrm{tr}(P_+ \theta^i_{10} 
(\theta^{i'}_{10})^\dagger) \,
\mathrm{tr}(Q_i^{10} (Q_{i'}^{10})^\dagger)  
+ \sum_{j,j'} \mathrm{tr}(P_+ \theta^j_{120} 
(\theta^{j'}_{120})^\dagger) \,
\mathrm{tr}(Q_j^{120} (Q_{j'}^{120})^\dagger) \big)\,. 
\end{align*}
The other parts of the Lagrangian are  
\[
\mathcal{L}_2 = \tfrac{\epsilon}{192 g^2} \mathrm{tr}((\mathbf{d}A 
+ A^2|_{\Lambda^2})^2)~,\qquad 
\mathcal{L}_1 = \tfrac{\epsilon}{192 g^2}
\mathrm{tr}((\mathbf{d}\pi(\eta) + \{A,\pi(\eta)-\iu Y\})^2)~.
\]

As there are $23$ different $\boldsymbol{1}$-terms, 
$48-2=46$ different $\boldsymbol{45}$-terms, 
$70-6=64$ different $\boldsymbol{210}$-terms, 
$6$ different $\boldsymbol{10}$-terms, 
$9$ different $\boldsymbol{120}$-terms and 
$7$ different $\boldsymbol{10}$-terms in $\theta$,  
there occur
$\tfrac{1}{2} (23 \cdot 24 + 46 \cdot 47 + 64 \cdot 65 + 6 \cdot 7 
+ 9 \cdot 10 + 7 \cdot 8) = 3531$
different gauge invariant terms in the Higgs potential. All of them are 
compatible with the configuration $\mathcal{M}$ and $\mathcal{N}$ specified 
by the Yukawa operator $Y$ as the vacuum. This means that the 
most general gauge invariant Higgs potential that leads to the desired 
spontaneous symmetry breaking depends on $3531$ parameters. Of course, not 
all these terms are really necessary, and in the classical formulation one 
puts most of the coefficients equal to zero. But there is no justification 
for doing so, the Higgs potential in the classical formulation does contain 
$3531$ parameters. Our theory reduces this huge number to $9$, namely the 
parameters of the unknown matrix $M_1$. All other matrices occur in the 
fermionic action and can be measured, in principle. Thus, although our 
$\mathrm{SO(10)}$ model has more independent parameters than the standard 
model, the ratio of the fixed parameters to the classical parameters is 
much better than in corresponding treatments of the standard model.

We now study the Yang-Mills part $\mathcal{L}_2 - \tfrac{\epsilon}{192 g^2} 
\mathrm{tr}(\{A,Y\}^2)$ of the Lagrangian. For that purpose we
introduce chiral $\Gamma$ matrices
\begin{align*}
\Gamma^\pm_1 &= \tfrac{1}{\sqrt{2}}(\Gamma_4 \pm \iu \Gamma_5)\,, &
\Gamma^\pm_2 &= \tfrac{1}{\sqrt{2}}(\Gamma_6 \pm \iu \Gamma_9)\,, &
\Gamma^\pm_3 &= \tfrac{1}{\sqrt{2}}(\Gamma_7 \pm \iu \Gamma_8)\,, \\
\Gamma^\pm_4 &= \tfrac{1}{\sqrt{2}}(\Gamma_0 \pm \iu \Gamma_3)\,, &
\Gamma^\pm_5 &= \tfrac{1}{\sqrt{2}}(\Gamma_1 \pm \iu \Gamma_2)\,, \\
\{\Gamma^p_i,\Gamma^q_j\} &= 2 \delta^{p\bar{q}}_{ij} \one_{32}\,, &
\delta^{p\bar{q}}_{ij}  &= \delta_{ij} \delta^{p\bar{q}} & 
\Gamma^{pq}_{ij} &= \tfrac{1}{2} [\Gamma^p_i,\Gamma^q_j] =
- (\Gamma^{\bar{p}\bar{q}}_{ij})^\dagger \,, 
\end{align*}
where $p,q\in \{+,-\}$ and $i,j= 1,\dots,5$. The bar in $\bar{q}$
changes the sign. The Yang-Mills field $A$ can now be decomposed as
\begin{align}
A &=\mathcal{P} (\tfrac{1}{4 \sqrt{2}} g A^{ij}_{pq,\mu} \gamma^\mu \otimes 
\Gamma^{pq}_{ij} \otimes \one_3 \otimes \one_4)
=\mathcal{P} (\gamma^\mu \otimes \mathbb{A}_\mu \otimes \one_3 
\otimes \one_4)~, \label{Aexpl} 
\\
\mathbb{A}_\mu &= \tfrac{1}{4\sqrt{2}} 
A^{ij}_{pq,\mu} \Gamma^{pq}_{ij}~, \qquad 
A^{ij}_{pq,\mu} = {-} A^{ji}_{qp,\mu} =
\overline{A^{ij}_{\bar{p}\bar{q},\mu}}  \in C^\infty(M)~, \notag
\end{align}
where $\one_3$ acts on the generation space and $\one_4$ is the 
$4\times 4$ matrix structure in \eqref{A}. Defining 
$\gamma^{\mu\nu}=\tfrac{1}{2} (\gamma^\mu \gamma^\nu - \gamma^\nu 
\gamma^\mu)$ we have 
\begin{align*}
\mathbf{d} A &=\mathcal{P}(\tfrac{1}{8 \sqrt{2}} g 
(\partial_\mu A^{ij}_{pq,\nu} - \partial_\nu A^{ij}_{pq,\mu}) 
\gamma^{\mu\nu} \otimes \Gamma^{pq}_{ij} \otimes \one_3 \otimes \one_4)~,
\\
A^2 &= \mathcal{P}(\tfrac{1}{32} g^2 A^{ij}_{pq,\mu} A^{kl}_{rs,\nu} 
\gamma^\mu \gamma^\nu \otimes \Gamma^{pq}_{ij} \Gamma^{rs}_{kl} 
\otimes \one_3 \otimes \one_4)
\\ &
= \mathcal{P}(\tfrac{1}{128} g^2 A^{ij}_{pq,\mu} A^{kl}_{pq,\nu} 
(\gamma^\mu \gamma^\nu + \gamma^\nu \gamma^\mu) \otimes 
(\Gamma^{pq}_{ij} \Gamma^{rs}_{kl} + \Gamma^{rs}_{kl} \Gamma^{pq}_{ij}) 
\otimes \one_3 \otimes \one_4)
\\ &
+ \mathcal{P}(\tfrac{1}{128} g^2 A^{ij}_{pq,\mu} A^{kl}_{pq,\nu} 
(\gamma^\mu \gamma^\nu - \gamma^\nu \gamma^\mu) \otimes 
(\Gamma^{pq}_{ij} \Gamma^{rs}_{kl} - \Gamma^{rs}_{kl} \Gamma^{pq}_{ij}) 
\otimes \one_3 \otimes \one_4)~.
\end{align*}
This gives with 
\begin{gather*}
(\Gamma^{pq}_{ij} \Gamma^{rs}_{kl}- \Gamma^{rs}_{kl} \Gamma^{pq}_{ij}) = 
2\, f^{mn,pqrs}_{tu,ijkl}\, \Gamma^{tu}_{mn}~, \\
f^{mn,pqrs}_{tu,ijkl}= 
\delta^{q\bar{r}}_{jk} \delta^{m,p}_{t,i}\delta^{n,s}_{u,l} -
\delta^{p\bar{r}}_{ik} \delta^{m,q}_{t,j}\delta^{n,s}_{u,l} -
\delta^{q\bar{s}}_{jl} \delta^{m,p}_{t,i}\delta^{n,r}_{u,k} +
\delta^{p\bar{s}}_{il} \delta^{m,q}_{t,j}\delta^{n,r}_{u,k} ~,
\end{gather*}
where $\delta^{i,q}_{p,j}=\delta^i_j \delta^q_p$, 
the final results
\begin{align*}
A^2|_{\Lambda^2} &= \mathcal{P}(\tfrac{1}{32} g^2 f^{mn,pqrs}_{tu,ijkl} 
A^{ij}_{pq,\mu} A^{kl}_{rs,\nu} \gamma^{\mu\nu} \otimes \Gamma^{tu}_{mn} 
\otimes \one_3 \otimes \one_4)~,
\\
\mathbf{d} A + A^2|_{\Lambda^2} 
&= \mathcal{P}( \tfrac{1}{8 \sqrt{2}} g F^{mn}_{tu,\mu\nu} 
\gamma^{\mu\nu} \otimes \Gamma^{tu}_{mn} \otimes \one_3 \otimes \one_4)~,
\\
F^{mn}_{tu,\mu\nu} &= \partial_\mu A^{mn}_{tu,\nu} 
- \partial_\nu A^{mn}_{tu,\mu} 
+ \tfrac{1}{2 \sqrt{2}} g f^{mn,pqrs}_{tu,ijkl} 
A^{ij}_{pq,\mu} A^{kl}_{rs,\nu} ~.
\end{align*}
Using $\mathrm{tr}(\gamma^{\mu\nu}\gamma^{\kappa\lambda})
{=} 4 (g^{\mu\lambda} g^{\nu\kappa} {-} g^{\mu\kappa} g^{\nu\lambda})$ and 
$\mathrm{tr}(\Gamma_{ij}^{pq}\Gamma_{kl}^{rs})
{=} 32 (\delta_{il}^{p\bar{s}}\delta_{jk}^{q\bar{r}} {-} 
\delta_{ik}^{p\bar{r}}\delta_{jl}^{q\bar{s}})$, we arrive at
\begin{align*}
\mathcal{L}_2 &= \tfrac{\epsilon}{192 g^2} \cdot \tfrac{1}{128} g^2 
F^{ij}_{pq,\mu\nu}F^{kl}_{rs,\kappa\lambda} \cdot
\mathrm{tr}(\gamma^{\mu\nu}\gamma^{\kappa\lambda})\cdot
\mathrm{tr}((P_+{+}P_-)\Gamma^{ij}_{pq}\Gamma^{kl}_{rs}) \cdot 6 
= \tfrac{\epsilon}{8} F_{pq,\mu\nu}^{ij} 
F^{\bar{p}\bar{q},\mu\nu}_{ij} . 
\end{align*}

It is now convenient to identify the $\mathrm{su(3)} \oplus
\mathrm{su(2)}_L \oplus \mathrm{su(2)}_R \oplus \mathrm{u(1)}_{B-L}$ 
gauge fields $(G^k,V^i,\tilde{V}^i,G^0)$, where from now on 
$i,j=1,2,3$ and $k=1,\dots,8$:
\begin{align*}
& A^{12}_{+-,\mu} = {-} \tfrac{1}{\sqrt{2}} \iu (G^6_\mu {-} \iu
G^7_\mu)\,, \quad
A^{23}_{+-,\mu} = {-}\tfrac{1}{\sqrt{2}} (G^4_\mu {+} \iu G^5_\mu)\,,
\quad
A^{31}_{+-,\mu} = {-}\tfrac{1}{\sqrt{2}} (G^1_\mu {-}\iu G^2_\mu)\,, 
\\ 
& \tfrac{1}{\sqrt{2}} (A^{33}_{+-,\mu} {-} A^{11}_{+-,\mu}) 
= \iu G^3_\mu\,, \qquad
\tfrac{1}{\sqrt{6}} (A^{33}_{+-,\mu} {+} A^{11}_{+-,\mu} 
{-} 2 A^{22}_{+-,\mu}) = \iu G^8_\mu\,, 
\\
& \tfrac{1}{\sqrt{3}} (A^{11}_{+-,\mu} {+} A^{22}_{+-,\mu} 
{+} A^{33}_{+-,\mu}) = \iu G^0_\mu \,, 
\\
& A^{45}_{-+,\mu} = -\tfrac{1}{\sqrt{2}} (V^1_\mu {-} \iu V^2_\mu) 
=: -V^+_\mu = -(V^-_\mu)^\dagger \,, \qquad
\tfrac{1}{\sqrt{2}} (A^{44}_{+-,\mu} {-} A^{55}_{+-,\mu}) 
= {-}\iu V^3_\mu\,, 
\\
& A^{45}_{++,\mu} = \tfrac{1}{\sqrt{2}} (\tilde{V}^1_\mu 
{-} \iu \tilde{V}^2_\mu) =: \tilde{V}^+_\mu = (\tilde{V}^-_\mu)^\dagger
\,, \qquad \tfrac{1}{\sqrt{2}} (A^{44}_{+-,\mu} {+} A^{55}_{+-,\mu}) 
= \iu \tilde{V}^3_\mu\,. 
\end{align*} 
in this notation, the Yang-Mills Lagrangian $\mathcal{L}_2$ takes the form
\begin{align}
\mathcal{L}_2 &= \tfrac{\epsilon}{4} \big(
G^k_{\mu\nu}G_k^{\mu\nu} + \partial_{[\mu} G^0_{\nu]} 
\partial^{[\mu} G_0^{\nu]} + V^i_{\mu\nu} V_i^{\mu\nu} 
+ \tilde{V}^i_{\mu\nu} \tilde{V}_i^{\mu\nu} \big) 
\\ &
\textstyle + \tfrac{\epsilon}{2} (\begin{array}[t]{l}
\partial_{[\mu} A^{ij}_{++,\nu]} \partial^{[\mu} A_{ij}^{--,\nu]} 
+ \partial_{[\mu} A^{i4}_{++,\nu]} \partial^{[\mu} A_{i4}^{--,\nu]} 
+ \partial_{[\mu} A^{i4}_{-+,\nu]} \partial^{[\mu} A_{i4}^{+-,\nu]} \\
+ \partial_{[\mu} A^{i5}_{++,\nu]} \partial^{[\mu} A_{i5}^{--,\nu]} 
+ \partial_{[\mu} A^{i5}_{-+,\nu]} \partial^{[\mu} A_{i5}^{+-,\nu]} )
+ I.T~, \end{array} \notag
\end{align}
where $I.T$ stands for interaction terms between 3 or 4 Yang-Mills
fields and
\begin{align*}
G^k_{\mu\nu} &= \partial_{[\mu} G^k_{\nu]} 
- g f^k_{k'k''} G^{k'}_\mu G^{k''}_\nu~, \\
V^i_{\mu\nu} &= \partial_{[\mu} V^i_{\nu]} - g \epsilon^i_{jj'} V^j_\mu
V^{j'}_\nu~, & 
\tilde{V}^i_{\mu\nu} &= \partial_{[\mu} \tilde{V}^i_{\nu]} 
- g \epsilon^i_{jj'} \tilde{V}^j_\mu \tilde{V}^{j'}_\nu~, 
\end{align*}
and $f^k_{k'k''}$ and $\epsilon^i_{jj'}$ are $\mathrm{su(3)}$ and 
$\mathrm{su(2)}$ structure constants. The lesson is that the choice
$\tfrac{1}{192 g^2}$ for the global normalization constant was
correct, where $g$ is the coupling constant of $\mathrm{su(3)}$ and
the two $\mathrm{su(2)}$ Lie subalgebras. 

It remains to compute the mass terms 
{\arraycolsep 0pt
\begin{align*}
-\epsilon & \tfrac{1}{192 g^2} \mathrm{tr}(\{A, Y\}^2) \\
&= - \epsilon \tfrac{1}{192 g^2} \cdot 4 \cdot \mathrm{tr}(
\gamma^\mu \gamma^5{}^* \gamma^\nu \gamma^5 \otimes
(P_+[\mathbb{A}_\mu,\mathcal{M}][\mathbb{A}_\nu,\mathcal{M}] +
P_+[\mathbb{A}_\mu,\mathcal{N}][\mathbb{A}_\nu,\mathcal{N}^\dagger]))
\\
&= \tfrac{1}{12 g^2} \cdot (\tfrac{g}{4 \sqrt{2}})^2 \cdot 4 \cdot 16
\cdot \big(
\\
& 4\,(A^{i4}_{-+,\mu} A_{i4}^{+-,\mu} + A^{i4}_{++,\mu}
A_{i4}^{--,\mu} )\, 
\mathrm{tr}(\begin{array}[t]{l} 2 M_1^2 + M_a' M_a'{}^\dagger 
+ M_a M_a^\dagger + 3 M_b' M_b'{}^\dagger + 3 M_b M_b^\dagger \\
+ 3 M_c M_c^\dagger + 3 M_f M_f^\dagger 
+ M_p M_p^\dagger + M_s M_s^\dagger) \end{array}
\\ 
& + 8\,(A^{i5}_{-+,\mu} A_{i5}^{+-,\mu} + A^{i5}_{++,\mu}
A_{i5}^{--,\mu} )\, 
\mathrm{tr}(\begin{array}[t]{l} M_1^2 + M_a' M_a'{}^\dagger 
+ M_a M_a^\dagger + M_b' M_b'{}^\dagger \\
+ M_b M_b^\dagger + 2 M_c M_c^\dagger + 2 M_f M_f^\dagger ) \end{array}
\\ 
&+ 2\, (A^{i4}_{+-,\mu} A_{i4}^{-+,\mu} + A^{i5}_{++,\mu}
A_{i5}^{--,\mu} )\, 
\mathrm{tr}(M_2 M_2^\dagger)
\\
& + 4\,(A^{i4}_{++,\mu} A_{i4}^{-+,\mu} + A^{i4}_{--,\mu}
A_{i4}^{+-,\mu}) \, 
\mathrm{tr}(\begin{array}[t]{l} M_a' M_a'{}^\dagger - M_a M_a^\dagger 
+ M_b' M_b'{}^\dagger - M_b M_b^\dagger \\
+ M_c M_c^\dagger - M_f M_f^\dagger 
+ M_s M_s^\dagger - M_p M_p^\dagger) \end{array}
\\  
& + 4\, (-A^{i4}_{++,\mu} A_{i4}^{-+,\mu} + A^{i4}_{--,\mu}
A_{i4}^{+-,\mu}) \, 
\mathrm{tr}(\begin{array}[t]{l} M_a' M_a^\dagger 
- M_a M_a'{}^\dagger + M_b' M_b^\dagger - M_b M_b'{}^\dagger \\ 
+ M_c M_f^\dagger - M_f M_c^\dagger + M_s M_p^\dagger 
- M_p M_s^\dagger) \end{array}
\\  
& + 8\, (A^{i5}_{++,\mu} A_{i5}^{--,\mu} -  A^{i5}_{+-,\mu}
A_{i5}^{-+,\mu} )\,  
\mathrm{tr}(\begin{array}[t]{l} M_b M_a^\dagger 
+ M_a M_b^\dagger + M_a' M_b'{}^\dagger + M_b' M_a'{}^\dagger \\
+ 2 M_c M_f^\dagger + 2 M_f M_c^\dagger) \end{array}
\\ 
&+ 32\, A^{ij}_{++,\mu} A_{ij}^{--,\mu} \, 
\mathrm{tr}( \begin{array}[t]{l} 
M_1^2 + M_b' M_b'{}^\dagger + M_b M_b^\dagger 
+ M_c M_c^\dagger + M_f M_f^\dagger 
+ \tfrac{1}{16} M_2 M_2^\dagger) \end{array}
\\
& + 8 \iu \epsilon^i_{jj'} \,A^{jj'}_{++,\mu} A_{i5}^{++,\mu} \,
\mathrm{tr}(M_2 M_f^\dagger {+} M_2 M_c^\dagger) 
- 8 \iu \epsilon^i_{jj'} \, A^{jj'}_{--,\mu} A_{i5}^{--,\mu} \, 
\mathrm{tr}(M_f M_2^\dagger {+} M_c M_2^\dagger )
\\
&+ 2\, (2 V^+_\mu V_-^\mu + 2 \tilde{V}^+_\mu \tilde{V}_-^\mu 
+ \begin{array}[t]{l} (V^3_\mu {-} \tilde{V}^3_\mu)
(V_3^\mu {-} \tilde{V}_3^\mu)) \, \mathrm{tr}( 
M_a' M_a'{}^\dagger + M_a M_a^\dagger + 3 M_b' M_b'{}^\dagger \\ 
+ 3 M_b M_b^\dagger + 3 M_c M_c^\dagger + 3 M_f M_f^\dagger 
+ M_s M_s^\dagger + M_p M_p^\dagger) \end{array}
\\
&- 4 (V^+_\mu \tilde{V}_-^\mu + V^-_\mu \tilde{V}_+^\mu ) \,
\mathrm{tr}( \begin{array}[t]{l} -M_a' M_a'{}^\dagger + M_a M_a^\dagger 
+ 3 M_b' M_b'{}^\dagger - 3 M_b M_b^\dagger \\ 
- 3 M_c M_c^\dagger + 3 M_f M_f^\dagger + M_s M_s^\dagger 
- M_p M_p^\dagger) \end{array}
\\
&- 4 (V^+_\mu \tilde{V}_-^\mu - V^-_\mu \tilde{V}_+^\mu ) \,
\mathrm{tr}( \begin{array}[t]{l} M_a M_a'{}^\dagger 
- M_a' M_a^\dagger + 3 M_b' M_b^\dagger - 3 M_b M_b'{}^\dagger \\ 
+ 3 M_f M_c^\dagger - 3 M_c M_f^\dagger 
+ M_s M_p^\dagger - M_p M_s^\dagger) 
\end{array}
\\
&+ 2 (\tilde{V}^+_\mu \tilde{V}_-^\mu 
+ (\sqrt{\tfrac{3}{2}} G^0_\mu {+} \tilde{V}^3_\mu)
(\sqrt{\tfrac{3}{2}} G_0^\mu {+} \tilde{V}_3^\mu))\, 
\mathrm{tr}(M_2 M_2^\dagger) 
\big)~,
\end{align*}}%
where we used $\mathrm{tr}(\gamma^\mu\gamma^\nu) =4 g^{\mu\nu}$ and 
$\gamma^5 \gamma^5{}^*=\epsilon \one_4$. The factor $4$ comes from the
anti-symmetry $A^{ij}_{pq}=-A^{ji}_{qp}$ and the $16$ from the trace over 
$P_+$ times $\Gamma$ matrices. 

We anticipate (section \ref{fermi}) the relation between fermion
masses and the matrices $M_{s,p,a',a,b',b,c,f}$, which reads ($tp=$
transpose of the preceding term)
\begin{align}
M_s  &= \tfrac{1}{16}(M_n {+} 3 M_u {+} M_e {+} 3 M_d) + tp~, &
M_a  &= \tfrac{1}{16}(M_n {+} 3 M_u {+} M_e {+} 3 M_d) - tp~, \notag \\
M_p  &= \tfrac{1}{16}(M_n {+} 3 M_u {-} M_e {-} 3 M_d) + tp~, &
M_a' &= \tfrac{1}{16}(M_n {+} 3 M_u {-} M_e {-} 3 M_d) - tp~, \notag \\
M_c  &= \tfrac{1}{16}(M_n {-}   M_u {-} M_e {+}   M_d) + tp~, &
M_b  &= \tfrac{1}{16}(M_n {-}   M_u {-} M_e {+}   M_d) - tp~, \notag \\
M_f  &= \tfrac{1}{16}(M_n {-}   M_u {+} M_e {-}   M_d) + tp~, &
M_b' &= \tfrac{1}{16}(M_n {-}   M_u {+} M_e {-}   M_d) - tp~.
\label{mm}
\end{align}
This gives
\begin{align*}
\tfrac{2}{3} \mathrm{tr} &(M_a' M_a'{}^\dagger {+} M_a M_a^\dagger 
{+} 3 M_b' M_b'{}^\dagger {+} 3 M_b M_b^\dagger {+} 3 M_c M_c^\dagger 
{+} 3 M_f M_f^\dagger 
{+} M_s M_s^\dagger {+} M_p M_p^\dagger) \\ &
= \tfrac{1}{12} \mathrm{tr}(M_n M_n^\dagger + M_e M_e^\dagger 
+ 3 M_u M_u^\dagger + 3 M_d M_d^\dagger) = \mu~,
\\
\tfrac{4}{3} \mathrm{tr} &({-} M_a' M_a'{}^\dagger {+} M_a M_a^\dagger 
{+} 3 M_b' M_b'{}^\dagger {-} 3 M_b M_b^\dagger {-} 3 M_c M_c^\dagger 
{+} 3 M_f M_f^\dagger {+} M_s M_s^\dagger {-} M_p M_p^\dagger) \\ &
= \tfrac{1}{6} \mathrm{tr}(M_n M_e^\dagger + M_e M_n^\dagger 
+3 M_u M_d^\dagger + 3 M_d M_u^\dagger) 
= 2 \tilde{\mu}~,
\\
\tfrac{4}{3} \mathrm{tr}&( M_a M_a'{}^\dagger {-} M_a' M_a^\dagger 
{+} 3 M_b' M_b^\dagger {-} 3 M_b M_b'{}^\dagger 
{+} 3 M_f M_c^\dagger {-} 3 M_c M_f^\dagger 
{+} M_s M_p^\dagger {-} M_p M_s^\dagger) \\ &
= \tfrac{1}{6} \mathrm{tr}(M_e M_n^\dagger {-} M_n M_e^\dagger 
{+} 3 M_d M_u^\dagger {-} 3 M_u M_d^\dagger) = 2 \iu \hat{\mu}\;.
\end{align*}
The numbers $\mu,\tilde{\mu},\hat{\mu}$ are not determined by the
experimental data because the Dirac mass matrix for the neutrinos
$M_n$ is unknown. Let us assume that the largest eigenvalue of $M_n$
is smaller than the mass $m_t$ of the top quark. Then, 
we have in leading approximation $\mu = \tfrac{1}{4} m_t^2$ and 
$\tilde{\mu}=\tfrac{1}{4} m_t m_b$. 

We now diagonalize the $V$-$\tilde{V}$-$G^0$ sector. The photon is the
massless linear combination $P_\mu = \tfrac{1}{2} G^0_\mu 
{-} \sqrt{\tfrac{3}{8}} V^3_\mu {-} \sqrt{\tfrac{3}{8}} \tilde{V}^3_\mu$,
which is perpendicular to the plane spanned by 
$\tfrac{1}{\sqrt{2}}(\tilde{V}^3_\mu {-} V^3_\mu)$ and 
$\sqrt{\tfrac{2}{5}} \tilde{V}^3_\mu {+} \sqrt{\tfrac{3}{5}}
G^0_\mu$. Now, abbreviating $M'{}^2 {=} \tfrac{1}{3} \mathrm{tr} 
(M_2 M_2^\dagger)$, the mass terms of the $V$-$\tilde{V}$-$G^0$ sector
are 
\begin{align*}
&\tfrac{1}{2} \Big(\tfrac{1}{\sqrt{2}}(\tilde{V}^3_\mu {-} V^3_\mu),\, 
\tfrac{1}{\sqrt{8}} (\tilde{V}^3_\mu {+} V^3_\mu) 
{+} \tfrac{\sqrt{3}}{2} G^0_\mu\Big) \!\!
\left( \!\! \begin{array}{cc} 2 \mu + M'{}^2 & 
2 M'{}^2 \\ 2 M'{}^2 & 4 M'{}^2 \end{array} \!\! \right) \!\!
\left(\!\!\!\begin{array}{c}  
\tfrac{1}{\sqrt{2}}(\tilde{V}^3_\mu {-} V^3_\mu) \\ 
\tfrac{1}{\sqrt{8}} (\tilde{V}^3_\mu {+} V^3_\mu) 
{+} \tfrac{\sqrt{3}}{2} G^0_\mu \end{array}\!\!\! \right) \\[-1ex]
& +
\Big( V^+_\mu \,,\, \tilde{V}^+_\mu \Big) \!\!
\left( \begin{array}{cc} \mu & -\tilde{\mu} - \iu \hat{\mu} \\
-\tilde{\mu} + \iu \hat{\mu} & \mu + M'{}^2 \end{array} \right) \!\!
\left(\begin{array}{c} V^-_\mu \\ \tilde{V}^-_\mu \end{array} \right)\,.
\end{align*}
After a unitary-orthogonal transformation
{\small
\begin{gather*}
\left( \!\!\! \begin{array}{c} Z_\mu \\ \tilde{Z}_\mu 
\end{array} \!\!\! \right)
\! = \! \left(\!\!\! \begin{array}{ccc} \cos \phi\!\! & - \sin \phi  \\ 
\sin \phi \!\! & \cos \phi \end{array}\!\!\! \right) \!\!
\left( \!\!\! \begin{array}{c}  
\tfrac{1}{\sqrt{2}}(V^3_\mu {+}\tilde{V}^3_\mu) 
\\ \tfrac{1}{\sqrt{8}} (\tilde{V}^3_\mu {-} V^3_\mu) 
{+} \tfrac{\sqrt{3}}{2} G^0_\mu \end{array} \!\!\! \right) ,~\;
\\ 
\left( \!\!\! \begin{array}{c} W^-_\mu \\ \tilde{W}^-_\mu 
\end{array} \!\!\! \right) \!
= \! \left( \!\!\! \begin{array}{cc} 
\cos \chi\!\! & {-} \mathrm{e}^{\iu \chi'} \sin \chi \\ \sin \chi \!\!
& \mathrm{e}^{\iu \chi'} \cos \chi \end{array} \!\!\! \right) \!\! 
\left( \!\!\! \begin{array}{c} V^-_\mu \\ \tilde{V}^-_\mu 
\end{array} \!\!\! \right) ,
\end{gather*}}%
where $\mathrm{e}^{\iu \chi'} = \tfrac{\tilde{\mu} + \iu \hat{\mu}}{
\sqrt{\hat{\mu}^2 + \tilde{\mu}^2}}$, 
the physical particles obtain the following masses:
\[
\begin{array}{lll}
m_W^2 && \mu + \tfrac{1}{2} M'{}^2 - \sqrt{\tfrac{1}{4} M'{}^4 
{+} \hat{\mu}^2 {+} \tilde{\mu}^2} \approx \mu + (\hat{\mu}^2 {+}
\tilde{\mu}^2)/M'{}^2  
\\
m_{\tilde{W}}^2 && \mu + \tfrac{1}{2} M'{}^2 + \sqrt{\tfrac{1}{4} M'{}^4 {+} 
\hat{\mu}^2 {+} \tilde{\mu}^2} \approx M'{}^2 + \mu  
\\
m_Z^2 && \mu + \tfrac{5}{2} M'{}^2 - \sqrt{ \tfrac{25}{4} M'{}^4
- 3 \mu M'{}^2 + \mu^2} \approx \tfrac{8}{5} \mu 
- \tfrac{16\,\mu^2}{25\,M'{}^2}
\\
m_{\tilde{Z}}^2 && \mu + \tfrac{5}{2} M'{}^2 + \sqrt{ \tfrac{25}{4} M'{}^4 
- 3 \mu M'{}^2 + \mu^2} \approx 5 M'{}^2 + \tfrac{2}{5} \mu 
\end{array}
\]
This means
\begin{align}
m_W &\approx \tfrac{1}{2} m_t~, & m_Z &= m_W/\cos \theta_W~, &
\sin^2 \theta_W &\approx \tfrac{3}{8} - \tfrac{m_t^2}{80\, M'{}^2}~.
\end{align}
We recall that the mass prediction for $m_W$ depends crucially on the
assumption that the Dirac mass for the neutrinos can be neglected. But
as $m_W$ cannot be much larger than $\tfrac{1}{2} m_t$, this
assumption seems to be correct. 
For the rotation angles we get $\cot 2\phi = \tfrac{3}{4} 
- \tfrac{\mu}{2 M'{}^2}$ and $\tan 2 \chi = {-} 2 \sqrt{\tilde{\mu}^2 
+ \hat{\mu}^2}/M'{}^2$. Hence, there is a violation of the standard model
of the order $m_t^2/M'{}^2$, for instance a coupling of the $W^\pm$
bosons to the right-handed fermions, and the Weinberg angle is not
universal any more. 

We neglect the mixing of the order $\|M_{a,a',b,b',c,f,s,p}\|/\|M_{1,2}\|$ 
between the very massive leptoquarks. Denoting $M^2=\tfrac{4}{3}
\mathrm{tr}(M_1^2)$, the masses of the leptoquarks are: 
\[
{\renewcommand{\arraystretch}{1.3}
\begin{array}{c|c|c|c|c}
A^{i4}_{+-} & A^{i4}_{++} & A^{i5}_{+-} & A^{i5}_{++} & A^{ij}_{++} 
\\ \hline
\sqrt{M^2 {+} M'{}^2} & M & M &
\sqrt{M^2 {+} M'{}^2} & \sqrt{4 M^2 {+} M'{}^2} 
\end{array}}
\]
The renormalization group analysis \cite{m} suggests 
$M \approx 10^{16}\mathrm{GeV}$. Below that energy, the original gauge
group $\mathrm{SO(10)}$ is broken to the intermediate symmetry group 
$\mathrm{SU(3)}_C {\times} \mathrm{SU(2)}_L {\times} \mathrm{SU(2)}_R
{\times} \mathrm{U(1)}_{B-L}$. At the scale
$M' \approx 10^9 \mathrm{GeV}$, the
subgroups $\mathrm{SU(2)}_R$ and $\mathrm{U(1)}_{B-L}$
are broken to the standard model symmetry group 
$\mathrm{SU(3)}_C {\times} \mathrm{SU(2)}_L {\times} \mathrm{U(1)}_{Y}$,
with the hyper-charge given by $\sqrt{\tfrac{3}{5}} \tilde{V}^3_\mu 
{-} \sqrt{\tfrac{2}{5}} G^0_\mu$. Finally, the fermion masses break
the standard model symmetry at the scale $m_t \approx 10^2 \mathrm{GeV}$ 
to the remaining symmetry $\mathrm{SU(3)}_C {\times} \mathrm{U(1)}_{EM}$. 
Hence, the only massless Yang-Mills fields are the photon $P_\mu$ and the 
eight gluons $G^a_\mu$. The Higgs mechanism consists in using the
other $45-9=36$ $\mathrm{SO(10)}$ gauge parameters to 
eliminate the 36 Goldstone bosons $[\pi(a),-\iu Y]$, which in turn
breaks the symmetry from $\mathrm{SO(10)}$ to the fermion symmetry 
$\mathrm{SU(3)}_C {\times} \mathrm{U(1)}_{EM}$. Thus, there are 
$45 + 2 {\cdot} 10 + 4 {\cdot} 120 
+ 3 {\cdot} 2 {\cdot} 126 - 36 = 1301-36=1265$ independent Higgs components.  

We compute now the upper limit for the mass of the standard model
Higgs field. It is obtained by evaluating the Higgs potential $V$ at
the configuration 
\begin{align}
\tilde{\Theta} &= (\Gamma_{45} + \Gamma_{78} + \Gamma_{69})~, & 
\notag \\
\tilde{\Upsilon}_1 &= (1+\phi) \Gamma_0 ~, &
\tilde{\Upsilon}_2 &= (1+\phi) \Gamma_3 ~, \notag
\\
\tilde{\Phi}_1 &= (1+\phi) \Gamma_{120} ~, &
\tilde{\Phi}_2 &= (1+\phi) \Gamma_{123} ~, \notag
\\
\tilde{\Phi}_3 &= (1+\phi) (\Gamma_{453} + \Gamma_{783} + \Gamma_{693}) ~, &
\tilde{\Phi}_4 &= (1+\phi) (\Gamma_{450} + \Gamma_{780} + \Gamma_{690}) ~, 
\notag \\
\tilde{\Psi}_1 &= (1+\phi)
(\Gamma_{01245} + \Gamma_{01278} + \Gamma_{01269}) ~,&
\tilde{\Psi}_2 &= (1+\phi)
(\Gamma_{31245} + \Gamma_{31278} + \Gamma_{31269}) ~, \notag
\\
\tilde{\Psi}_3 &= \tfrac{1}{8} 
(\Gamma_1 {-} \iu \Gamma_2) \Gamma_3 (\Gamma_4 {-} \iu \Gamma_5)
(\Gamma_6 {-} \iu \Gamma_9)(\Gamma_7 {-} \iu
\Gamma_8)~. \hspace*{-\textwidth} 
\label{phi}
\end{align}
It is important that $\phi$ is a real field, because the configuration 
corresponding to an imaginary part is the Goldstone boson given by 
the commutator with $\Gamma^{+-}_{44}$. One easily finds 
\begin{align*}
\theta_{(1)} &= -\phi \one_4 \otimes \one_{32} \otimes \!\!
\begin{array}[t]{l} (\tilde{M}_{\{ss\}} 
+ \tilde{M}_{\{pp\}} + \tilde{M}_{\{a'a'\}} + \tilde{M}_{\{aa\}} 
+ 3 \tilde{M}_{\{b'b'\}} + 3 \tilde{M}_{\{bb\}} \\ 
+ 3 \tilde{M}_{\{cc\}} + 3 \tilde{M}_{\{ff\}}) \end{array}
\\
& + 2 \phi \one_4 {\otimes} \iu (\Gamma_{45} {+} \Gamma_{78} {+} \Gamma_{69}) 
{\otimes} (M_{\{sb'\}} {+} M_{\{pb\}} {+} M_{\{a'c\}} {+} M_{\{af\}} 
{+} 2 M_{\{bc\}} {+} 2 M_{\{b'f\}}) \;,\\
\theta_{(2)} &= 0~, 
\end{align*}
up to the Higgs self-interaction $\phi^2$. The other 
$\boldsymbol{45}$- and all $\boldsymbol{210}$-contributions are 
canceled due to \eqref{cns}. We insert \eqref{mm} and arrive at
{\arraycolsep 0.1em
\begin{align*}
\theta_{(1)} = -\tfrac{1}{16} \phi \one_4 &\otimes \one_{32} \otimes 
(\begin{array}[t]{l} \tilde{M}_{\{nn\}} + \tilde{M}_{\{ee\}} 
+ 3 \tilde{M}_{\{uu\}} + 3 \tilde{M}_{\{dd\}} \\
+ \tilde{M}_{\{n^t n^t\}} + \tilde{M}_{\{e^t e^t\}} 
+ 3 \tilde{M}_{\{u^t u^t\}} + 3 \tilde{M}_{\{d^t d^t\}} ) \end{array} 
\\
+ \tfrac{1}{16} \phi \one_4 &\otimes \iu (\Gamma_{45} + \Gamma_{78} 
+ \Gamma_{69}) \otimes (\begin{array}[t]{l} M_{\{nn\}} + M_{\{ee\}} 
- M_{\{uu\}} - M_{\{dd\}} \\
- M_{\{n^t n^t\}} - M_{\{e^t e^t\}} + M_{\{u^t u^t\}} 
+ M_{\{d^t d^t\}} )~, \end{array}
\end{align*}}%
where $\tilde{M}_{\alpha^t\alpha^t}:=(2 M_\alpha^T M_\alpha^T{}^\dagger 
)_\perp$. We neglect again $M_n$ and choose 
$M_u=\mathrm{diag}(m_u,m_c,m_t)$, where the entries are the masses 
of the $u,c,t$ quarks. There is a negligible contribution of $\theta_{45}$ 
to the Higgs potential, and we have in leading approximation
\[
\theta_{(1)}= -\tfrac{1}{4} \phi m_t^2 \one_4 \otimes \one_{32} \otimes 
\mathrm{diag}(-1,-1,2)~.
\]
This leads to 
\[
V = \tfrac{\epsilon}{48 g^2} \tfrac{1}{16} \phi^2 m_t^4 \cdot 4 
\cdot 16 \cdot 6 = \epsilon \tfrac{1}{2 g^2} m_t^4 \phi^2 ~.
\]
Inserting the configuration \eqref{phi} into the part $\mathcal{L}_1$ of the 
Lagrangian we get
\begin{align*}
\mathcal{L}_1 &= \tfrac{\epsilon}{192 g^2} \cdot 4 \cdot \mathrm{tr}(P_+
(\gamma^\mu \partial_\mu \gamma^5 \eta_\mathcal{N})^*
(\gamma^\nu \partial_\nu \gamma^5 \eta_\mathcal{N}))
= \tfrac{1}{12 g^2} \mathrm{tr}(P_+ \partial_\mu \eta_\mathcal{N}{}^\dagger 
\, \partial^\mu \eta_\mathcal{N}) 
\\
&= \tfrac{1}{12 g^2} \partial_\nu \phi \,\partial^\nu \phi \cdot 16 \cdot 
\mathrm{tr}(M_a' M_a'{}^\dagger + M_a M_a^\dagger + 3 M_b' M_b'{}^\dagger 
+ 3 M_b M_b^\dagger \\
&\hspace*{10em}+ 3 M_c M_c^\dagger + 3 M_f M_f^\dagger 
+ M_s M_s^\dagger + M_p M_p^\dagger) 
\\
&= \tfrac{2}{g^2} \mu \,\partial_\nu \phi\, \partial^\nu \phi~.
\end{align*}
This means that the physical Higgs boson is obtained by rescaling
$\varphi=\tfrac{g}{2\sqrt{\mu}} \phi = g \phi/m_t$, 
and it receives the mass 
\begin{equation}
m_\varphi = m_t~.
\end{equation}
This value is an upper bound for the Higgs mass, because we have only 
calculated the 
diagonal matrix element of the whole mass matrix. Due to the off-diagonal 
matrix elements, the masses = eigenvalues are different from the diagonal 
matrix elements, and the smallest eigenvalue is smaller than the 
smallest diagonal matrix element. It is plausible that this smallest diagonal 
matrix element is just $m_\varphi^2$, because any other Higgs configuration 
obtains a mass contribution from $\theta_{(2)}$ of the 
order $\mathrm{tr}((M_1 M_i \pm M_i \overline{M}_1)^2)/\mathrm{tr}(M_i^2)$. 

The computation of the masses of the remaining 1264 Higgs bosons is
analogous.

\section{The fermionic action}
\label{fermi}

To write down the fermionic action (in Minkowski space!), recall that
our setting has two symmetries $J$ and $\mathcal{S}$. 
It is therefore natural to demand that the fermionic configuration
space has the same symmetries,
\[
\mathcal{H}=\{\boldsymbol{\psi}\in L^2(M,S) \otimes \mathbb{C}^{384}:~~
\mathcal{P}\boldsymbol{\psi}=J \boldsymbol{\psi} 
= \mathcal{S} \boldsymbol{\psi} =\boldsymbol{\psi}\}~.
\]
As usual we impose a Weyl condition in Minkowskian case. This means to 
look for a chirality operator that commutes with both $J$ and $\mathcal{S}$. 
The unique choice up to the sign is 
\[
\boldsymbol{\psi}=\boldsymbol{\chi}\boldsymbol{\psi}~,\quad
\boldsymbol{\chi}= \mathcal{P} \mathrm{diag}(
-\gamma^5 \otimes \one_{96}\,,\, -\gamma^5 \otimes \one_{96} \,,\,
\gamma^5 \otimes \one_{96} \,,\, \gamma^5 \otimes \one_{96}) \mathcal{P}~.
\]
Thus, elements of $\mathcal{H}$ are of the form 
\[
\boldsymbol{\psi}=\left( \begin{array}{c}
(\tfrac{1}{2} (\one_4 {-} \gamma^5) \otimes P_+) 
\tilde{\boldsymbol{\psi}} \\
(- \iu \tfrac{1}{2} (\one_4 {-} \gamma^5) \otimes P_+ )
\tilde{\boldsymbol{\psi}} \\
(\tfrac{1}{2} (\one_4 {+} \gamma^5) \gamma^2 \otimes P_- B )
\overline{\tilde{\boldsymbol{\psi}}} \\
(\iu  \tfrac{1}{2} (\one_4 {+} \gamma^5) \gamma^2 \otimes P_- B )
\overline{\tilde{\boldsymbol{\psi}}} 
\end{array} \right)~, \qquad \tilde{\boldsymbol{\psi}} \in L^2(M,S) \otimes 
\mathbb{C}^{96}~.
\]

In order to eliminate the projection operators we introduce 
\begin{align*}
&B=\left( \begin{array}{cc} 0 & b \\ b & 0 \end{array} \right)~,\quad 
b=\rho_2 \eta_2 \tau_2 \sigma_2 \otimes \one_3 \in 
\mathbb{C}^{48}~,
\\  
&\mathbb{A}_\mu \otimes \one_3 
= \left( \begin{array}{cc} \boldsymbol{A}_\mu & 0 \\ 
0 & -b \boldsymbol{A}_\mu^T b \end{array} \right)~,\quad 
\boldsymbol{A}_\mu \in C^\infty(M) \otimes \mathrm{so(10)} \otimes \one_3~,
\\
&P_+ (\eta_\mathcal{N} + \mathcal{N}) P_- 
= \left( \begin{array}{cc} 0 & \tilde{H} \\ 0 & 0 \end{array} \right)~,
\\
& \sigma^0=\tilde{\sigma}^0 = \one_2~,\quad \sigma^a=-\tilde{\sigma}^a~,~~
a=1,2,3~,
\\
& (\tfrac{1}{2} (\one_4 {-} \gamma^5) \otimes P_+) \tilde{\boldsymbol{\psi}} 
= \big( 0\,,~ \psi_L \,,~ 0 \,,~ 0 \big)^t~,\quad 
\psi_L \in L^2(M)\mathbb{C}^2 \otimes \mathbb{C}^{16} \otimes 
\mathbb{C}^3~.
\end{align*}
Now, the fermionic action \eqref{SF} is 
\begin{align}
S_F &= \int_M dx \; \tfrac{1}{4} \boldsymbol{\psi}^\dagger \gamma^0 
(D+\iu \rho) \boldsymbol{\psi} \notag \\
&= \int_M dx \; \tfrac{1}{2} \Big( \psi_L^\dagger \,,\, 
- \psi_L^T \sigma^2 b \Big) \!\! \left( \begin{array}{cc} 
\iu \tilde{\sigma}^\mu (\partial_\mu + \boldsymbol{A}_\mu) & \tilde{H} \\ 
\tilde{H}^\dagger & \iu \sigma^\mu (\partial_\mu - b \boldsymbol{A}^T_\mu b) 
\end{array} \right)\!\! \left(\!\!\! \begin{array}{c}  
\psi_L \\ - \sigma^2 b \overline{\psi_L} 
\end{array} \!\!\! \right) \notag
\\
&= \int_M dx \; \Big(\iu \psi_L^\dagger \tilde{\sigma}^\mu 
(\partial_\mu + \boldsymbol{A}_\mu) \psi_L 
+ \Big\{ \tfrac{1}{2} \psi_L^\dagger \tilde{H} 
(- \sigma^2 b \overline{\psi_L}) + h.c \Big\} \Big)~. \label{sf}
\end{align}
To get the last line one has to take into account that fermions $\psi_i$ are 
Grassmann-valued, which means $\psi_1^T X^T \overline{\psi_2} = 
- \psi_2^\dagger X \psi_1$, for any matrix $X$. The correct fermionic 
parameterization dictated by the electric charge is
\begin{align*}
\psi_L &= \big( 
n_L, u_L^1, u_L^2, u_L^3, e_L, d_L^1, d_L^2, d_L^3, \\ & \hspace*{6em}
\sigma^2 \overline{d_R^3}, -\sigma^2 \overline{d_R^2}, 
-\sigma^2 \overline{d_R^1}, \sigma^2 \overline{e_R}, 
-\sigma^2 \overline{u_R^3}, \sigma^2 \overline{u_R^2}, 
\sigma^2 \overline{u_R^1}, -\sigma^2 \overline{\nu_R} \big)^t,
\\
-\sigma_2 b \overline{\psi_L} &= \big( 
{-} n_R, -u_R^1, -u_R^2, -u_R^3, -e_R, -d_R^1, -d_R^2, -d_R^3, 
\\ & \hspace*{6em}
\sigma^2 \overline{d_L^3}, -\sigma^2 \overline{d_L^2}, 
-\sigma^2 \overline{d_L^1}, \sigma^2 \overline{e_L}, 
-\sigma^2 \overline{u_L^3}, \sigma^2 \overline{u_L^2}, 
\sigma^2 \overline{u_L^1}, -\sigma^2 \overline{\nu_L} \big)^t,
\end{align*}
where $^t$ does not transpose the entries of the vector. The fermion
masses are produced by the term 
\begin{align*}
\tfrac{1}{2} \psi_L^\dagger \mathcal{N} 
(- \sigma^2 b \overline{\psi_L}) + h.c 
= \big\{ &- u_L^\dagger (M_u \otimes \one_3) u_R 
- d_L^\dagger (M_d \otimes \one_3) d_R \\ &
- e_L^\dagger M_e e_R - \nu_L^\dagger M_\nu \nu_R 
- \nu_R^T \sigma_2 M_2 \nu_R \big\} + h.c 
\end{align*}
in \eqref{sf}, with the mass matrices $M_{u,d,\nu,e}$ given implicitly
in \eqref{mm}. The right neutrinos receive a large Majorana mass of
the order $\|M_2\|$ and the see-saw mechanism produces very small
masses for the left-handed neutrinos of the order $\|m_n^2\|/\|M_2\|$.

\section{Outlook}

What we have presented here is a maximal $\mathrm{SO(10)}$ model which
allows the fermion masses to be as general as possible. This is in
contrast to the original idea of grand unification, namely, to reduce
the number of free parameters of the standard model. 
The number of Higgs multiplets can be reduced by imposing appropriate 
relations between the fermion masses. For instance, the minimal 
$\mathrm{SO(10)}$ model containing one complex $\boldsymbol{10}$, one 
complex $\boldsymbol{126}$ and the $\boldsymbol{45}$ (or 
$\boldsymbol{210}$) is obtained by putting 
$M_s=\lambda_1 M_p$, $M_2=\lambda_2 M_c=\lambda_3 M_f$ and 
$M_a=M_a'=M_b=M_b'=0$, with real parameters $\lambda_i$. This model is
very predictive in the fermion sector and one can calculate the
neutrino masses \cite{m}. However, in our formulation the ideal
$\mathcal{J}^2$ becomes so large that the only surviving terms in the
Higgs potential are $\boldsymbol{1}$ and $\boldsymbol{10}$. This is
not sufficient. There seems to be a strong evidence that a
$\boldsymbol{120}$ representation must be included. This
next-to-minimal $\mathrm{SO(10)}$ model will be studied elsewhere.

\section*{Acknowledgment}

I am grateful to Bruno Iochum, Daniel Kastler, Thomas Krajewski, Serge
\mbox{Lazzarini} and Thomas Sch\"ucker for very useful discussions and
for the hospitality at the CPT Luminy.


\begin{thebibliography}{99}

\bibitem{acf} A.~Connes, \emph{G\'eom\'etrie non commutative},
InterEditions (Paris 1990).

\bibitem{acr} A.~Connes, \emph{Noncommutative geometry and reality},
J.\ Math.\ Phys.\ \textbf{36} (1995) 6194--6231.

\bibitem{acg} A.~Connes, \emph{Gravity coupled with matter and the
foundation of noncommutative geometry}, Commun.\ Math.\ Phys.\
\textbf{182} (1996) 155--176.
 
\bibitem{dk} D.~Kastler, \emph{Noncommutative geometry and fundamental
physical interactions (Historical sketch and outline of the
present situation)}, Monsaraz summer school (1997).

\bibitem{lmms} F.~Lizzi, G.~Mangano, G.~Miele and G.~Sparano,
\emph{Constraints on unified gauge theories from noncommutative
geometry}, Mod.\ Phys.\ Lett.\ A \textbf{11} (1996) 2561--2572.
 
\bibitem{rw2} R.~Wulkenhaar, \emph{Non-commutative geometry with graded
differential Lie algebras}, J.\ Math.\ Phys.\ \textbf{38} (1997)
3358--3390.

\bibitem{fm} H.~Fritzsch and P.~Minkowski, \emph{Unified interactions
of leptons and hadrons}, Annals Phys.\ \textbf{93} (1975) 193--266.

\bibitem{g} H. Georgi, \emph{The state of the art--gauge theories},
in: Proc.\ 1974 Williamsburg AIP conf., ed by C.~E.~Carlton, AIP (New
York 1975) 575--582.

\bibitem{cf} A.~H.~Chamseddine and J.~Fr\"ohlich,
\emph{$\mathrm{SO(10)}$ unification in noncommutative geometry},
Phys.\ Rev.\ D \textbf{50} (1994) 2893--2907.

\bibitem{m} R.~N.~Mohapatra, \emph{Minimal $\mathrm{SO(10)}$ grand
unification: Predictions for proton decay and neutrino masses and
mixings}, Warsaw Elem.\ Part.\ Phys.\ (1993) 158--172, and 
\texttt{hep-ph/9310265}.

\end{thebibliography}
\end{document}